\documentclass[aps,eqsecnum,superscriptaddress,nofootinbib,11pt]{revtex4-2}

\usepackage{graphicx} 
\usepackage{bm} 

\usepackage{amsmath} 
\usepackage{amssymb} 
\usepackage{bbm} 
\usepackage{footnotebackref} 

\newcommand{\abs}[1]{{\left|{#1}\right|}} 
\newcommand{\ket}[1]{\vert{#1}\rangle} 
\newcommand{\bra}[1]{\langle{#1}\vert} 

\newcommand{\secref}[1]{Sec.~\ref{#1}}
\newcommand{\figref}[1]{Fig.~\ref{#1}}

\newcommand{\appref}[1]{Appendix~\ref{#1}}

\begin{document}
\count\footins = 1000 

\title{Response of the Unruh-DeWitt detector in a gravitational wave background}


\author{Bo-Hung Chen}
\email{kenny81778189@gmail.com}
\affiliation{Department of Physics, National Taiwan University, Taipei 10617, Taiwan}

\author{Dah-Wei Chiou}
\email{dwchiou@gmail.com}
\affiliation{Department of Physics, National Sun Yat-sen University, Kaohsiung 80424, Taiwan}
\affiliation{Center for Condensed Matter Sciences, National Taiwan University, Taipei 10617, Taiwan}

\begin{abstract}
Applying the techniques of light-front quantization to quantize a scalar field in a monochromatic gravitational wave background, we manage to investigate the response of the Unruh-DeWitt detector coupled to a scalar field in the presence of a gravitational wave for the two cases moving along a free-falling trajectory and a constant-accelerating trajectory. The transition rate of the Unruh-DeWitt detector, in both cases, is different from the result with no gravitational wave, and the leading-order correction due to the gravitational wave survives the long-wavelength limit that formally takes the wavelength of the gravitational wave to infinity. This new effect of the gravitational wave on a quantum system is qualitatively different from that on a classical mechanical system, and cannot be understood in terms of gravitational wave tidal force.
\end{abstract}

\maketitle


\section{Motivations and overview}
One of the most vexed problems that causes considerable confusion about gravitational waves is what their physical effects on matter really are. The confusion arose partly because one cannot define a localized energy-momentum for the gravitational field, and consequently the notion of stress-energy tensor of gravitational waves makes sense only as an average over several wavelengths in a ``coarse-grain'' sense (see Sections 20.4, 35.7, and 35.15 in \cite{Misner:1974qy}), rendering it somewhat unclear how the energy of a gravitational wave is transferred into matter.
Opinions on this issue had widely diverged until the \emph{sticky bead argument} was proposed (anonymously) by Feynman \cite{Feynman:a,Feynman:b} and by Bondi \cite{Bondi:1957dt}. The sticky bead argument suggests that, as a gravitational wave passes over two beads sliding with friction on a rigid rod, the beads will rub against the rod, thus absorbing some of the energy carried by the wave and dissipating it into heat.
A similar argument also applies to different classical mechanical systems, the response of which can be used to detect gravitational waves (see Chapter 37 of \cite{Misner:1974qy}). A notable example is the \emph{resonant mass detector}, which has been operated in various experiments (see \cite{Aguiar:2010kn} for a review) as alternatives to interferometric gravitational wave detectors (see \cite{Saulson:2017gqp,Reitze:2019nwo} for reviews).

To derive the response of a classical mechanical system to a gravitational wave, the equation of motion for mass elements of the system is dealt with in a standard Newtonian manner, except that, as a non-Newtonian effect, the tidal force produced by the gravitational wave provides the driving force against the Newtonian interacting force (e.g., friction in the sticky bead system, elastic and damping forces in the resonant mass detector, etc.) between mass elements (see Chapter 37, especially Section 37.2, of \cite{Misner:1974qy} for a detailed account).
This analysis is straightforward and easy to understand, but it might not manifest some subtle effects of gravitational waves not directly resulting from the tidal force.
Therefore, instead of phenomenologically considering the response of a classical mechanical system, it will yield valuable new insight into the gravitational wave effects on matter, if the response of a quantum system to gravitational waves can be studied from a more fundamental setting.

In this paper, we consider probably the simplest kind of such a theoretical quantum system --- the Unruh-DeWitt detector coupled to a massless real scalar field, and manage to investigate its response to a gravitational wave background for the two cases of a free-falling trajectory and a constant-accelerating trajectory. Our investigation shows that, in both cases, the transition rate of the Unruh-DeWitt detector is different from the result with no gravitational wave, and the leading-order correction due to the gravitational wave survives the long-wavelength limit --- an intriguing effect that cannot be explained out in terms of the gravitational wave tidal force.\footnote{The underlying mechanism of state transition of the Unruh-DeWitt detector is that the detector is coupled to quantum fluctuations of a quantum field of interest in the vacuum background, akin to spontaneous emission of an atom as a consequence of being coupled to quantum fluctuations of the electromagnetic field. Accordingly, simply by moving the Unruh-DeWitt detector in the vacuum background, it will yield a certain transition rate, which is measurable at least in principle. Experimentally speaking, it seems more realistic to model the detector as coupled to the electromagnetic field, instead of a massless scalar field, as Nature evidently has quantum fluctuations of the former in vacuum, but may not of the latter. Nevertheless, we adopt the model of a real scalar field, because not only it is theoretically the simplest but also it may still give the same measurement result of the detector that is coupled to the electromagnetic field but insensitive to its polarization. \secref{sec:Unruh-DeWitt detector} will further elaborate on the related issues concerning measurement.}

In the literature of the Unruh-DeWitt detector, it has been shown that the response of the Unruh-DeWitt detector is modified in the presence of boundaries \cite{Chiou:2016exd,Davies:1989me}, essentially because the boundary condition alters the mode expansion of the quantum field. In the limit that the length scale delimited by the boundaries goes to infinity, the modification becomes negligible.
In the case of a gravitational wave background, the detector's response is anticipated to change as well, since the gravitational wave also alters the mode expansion. However, in the formal limit that the wavelength of the gravitational wave goes to infinity, this change does not diminish as long as the amplitude of the gravitational wave remains finite. The gravitational wave effect on the Unruh-DeWitt detector is more involved than merely imposing a large length scale of the wavelength.

This paper is organized as follows.\footnote{Throughout this paper, we adopt the convention $(-,+,+,+)$ for the metric signature and use the natural units with both the Plank constant $\hbar$ and the speed of light $c$ set to unity.}
In \secref{sec:Unruh-DeWitt detector}, we give a brief review on the Unruh-DeWitt detector.\footnote{\secref{sec:Unruh-DeWitt detector} is based on Sec.~II and Appendix A of \cite{Chiou:2016exd}.}
In \secref{sec:scalar field}, we solve the equation of motion of a scalar field, i.e., the Klein-Gordon
equation, in a gravitational wave background. We then quantize the scalar field in the gravitational wave background using the light-front quantization \cite{Burkardt:1995ct} in \secref{sec:LF quantization}, and compute the Wightman function in \secref{sec:Wightman function}. With the Wightman function at hand, we compute the response of the Unruh-DeWitt detector along a free-falling trajectory and a constant-accelerating trajectory in \secref{sec:free-falling} and \secref{sec:constant-accelerating}, respectively. Finally, in \secref{sec:summary}, the results and their implications are summarized and remarked.
Additionally, we also explicitly solve the geodesic equation in a gravitation wave background in \appref{app:geodesic eq}.

\section{The Unruh-DeWitt detector}\label{sec:Unruh-DeWitt detector}
Whereas the notion of ``particles'' of a quantum field is clear to recognize and understand in flat spacetime, it is rather ambiguous in the context of quantum field theory in curved spacetime, as the particle content, quite surprisingly, turns out to be observer-dependent \cite{Fulling:1972md}. To have an unequivocal notion of particles, it thus requires an operational definition in terms of the response of a well-defined ``particle detector''. (The idea of a particle detector has already been considered for a different motivation in quantum optics by Glauber in 1963 \cite{Glauber:1963fi}.)
In 1976, Unruh proposed a theoretical model of such a particle detector and used it to address the problem of the particle content in relation to the observer's trajectory \cite{Unruh:1976db}. Unruh's detector is modeled as a point object in a small box coupled to the quantum field of interest, by which a particle of the quantum field is said to be detected if the object in the box is excited from its initial ground state to some excited state. (A similar model was also developed by S\'{a}nchez in 1981 \cite{Sanchez:1981xx}.)
In 1979, DeWitt \cite{DeWitt:1979} further improved Unruh's idea by simplifying the model as a two-level point monopole detector, which is now generally referred to as the \emph{Unruh-DeWitt detector} and widely used as a theoretical tool to probe quantum field effects in various settings of spacetime.

In this section, we briefly review the model of the Unruh-DeWitt detector, following closely the line of Sec.\ 3.3 in \cite{Birrell:1982ix}. Unlike \cite{Birrell:1982ix}, we consider the transition rates of both excitation ($\Delta E>0$) and de-excitation ($\Delta E<0$), and also take into account the switching function $\chi(\tau)$ as introduced in \cite{Louko:2006zv,Satz:2006kb,Louko:2007mu} in order to address the issue of regularization. We also briefly recap some passages in Appendix A of \cite{Chiou:2016exd} to address the measurement of the transition rate and the concept of detailed balance. For more about the Unruh-DeWitt detector and also the Unruh effect, see \cite{Birrell:1982ix,Wald:book,Padmanabhan:2003gd} and especially the comprehensive review article \cite{Crispino:2007eb}.

\begin{figure}
\centering
    \includegraphics[]{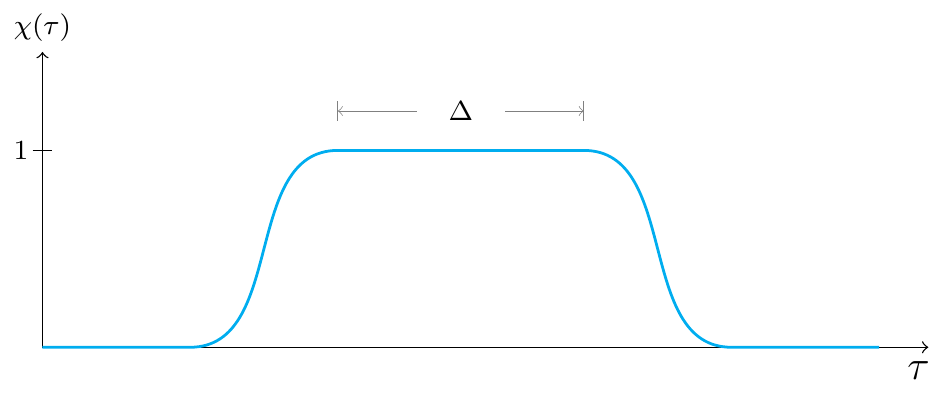}
\caption{A typical switching function $\chi(\tau)$ with a finite timescale $\Delta$ for the switch-on period.}
\label{fig:switching function}
\end{figure}

The Unruh-DeWitt detector is an idealized model with two energy levels, $\ket{E_0}$ and $\ket{E}$, coupled to a scalar field $\phi$ via a monopole interaction. If the detector moves along a world line $x^\mu(\tau)$, where $\tau$ is the detector's proper time, the Lagrangian for the monopole interaction is given by
\begin{equation}\label{monopole  interaction}
\kappa\,\chi(\tau)\mu(\tau)\phi(x^\mu(\tau)),
\end{equation}
where $\kappa$ is a small coupling constant, $\mu(\tau)$ is the operator of the detector's monopole moment, and $\chi(\tau)$ is the switching function, which accounts for the switch-on and switch-off of the detector. As the switching function $\chi(\tau)$ can be modeled as a smooth enough function with a compact support as depicted in \figref{fig:switching function}, its inclusion introduces a finite timescale $\Delta$ for the switch-on period.

Moving along a given trajectory $x^\mu(\tau)$, the detector in general does not remain in its initial state $\ket{E_0}$ but can be excited (if $\Delta E := E-E_0 >0$) or de-excited (if $\Delta E<0$) to the other state $\ket{E}$, while at the same time the field $\phi$ makes a transition from the vacuum state $\ket{0}$ to an excited state $\ket{\Psi}$. By the first-order perturbation theory, the amplitude for the transition
\begin{equation}\label{transition}
\ket{0,E_0}\rightarrow\ket{\Psi,E}
\end{equation}
is given by
\begin{equation}\label{transition amplitude 0}
i\kappa\,\bra{\Psi,E}\int_{-\infty}^\infty \chi(\tau) \mu(\tau)\, \phi\left(x^\mu(\tau)\right) d\tau \ket{0,E_0},
\end{equation}
which leads to the factorized form
\begin{equation}\label{transition amplitude}
i\kappa \bra{E}\mu(0)\ket{E_0} \int_{-\infty}^\infty e^{i(E-E_0)\tau} \chi(\tau) \bra{\Psi}\phi\left(x^\mu(\tau)\right)\ket{0}\,d\tau
\end{equation}
by the equation of evolution for $\mu(\tau)$,
\begin{equation}
\mu(\tau)=e^{iH_0\tau}\mu(0)e^{-iH_0\tau},
\end{equation}
where $H_0$ is the Hamiltonian of the detector.
Summing the squared norm of the amplitude given in \eqref{transition amplitude} over all possible $\ket{\Psi}$,\footnote{Here, we use the completeness relation $\sum_{\ket{\Psi}}\ket{\Psi}\bra{\Psi}=\mathbbm{1}$, but note that, at the level of the first-order perturbation, only the one-particle states of $\ket{\Psi}$ contribute.} we obtain the transition probability of $\ket{E_0}\rightarrow\ket{E}$ as
\begin{equation}\label{transition probability}
\kappa^2\abs{\bra{E}\mu(0)\ket{E_0}}^2\ F(E-E_0),
\end{equation}
where
\begin{equation}\label{response function}
F(\Delta E) = \int_{-\infty}^\infty d\tau \int_{-\infty}^\infty d\tau'
e^{-i\Delta E(\tau-\tau')} \chi(\tau)\,\chi(\tau')\, D^+(x(\tau),x(\tau'))
\end{equation}
is the \emph{response function}, which depends on the trajectory but not the internal properties of the detector. The remaining factor, $c^2\abs{\bra{E}\mu(0)\ket{E_0}}^2$, represents the \emph{selectivity}, which depends only on the detector's internal properties.\footnote{In following sections, we will focus on the response function $F(\Delta E)$ and ignore the factor of selectivity.} The Wightman functions $D^{\pm}$ are defined as
\begin{subequations}
\begin{eqnarray}
D^+(x,x')&:=&\bra{0}\phi(x)\phi(x')\ket{0},\\
D^-(x,x')&:=&\bra{0}\phi(x')\phi(x)\ket{0}.
\end{eqnarray}
\end{subequations}
It should be noted that the first-order perturbation method is viable only for a short range of evolution time, since the squared norm of transition amplitude has to remain much smaller than unity. Therefore, \eqref{transition amplitude 0} with the unbounded integral $\int_{\infty}^\infty d\tau$ is problematic, unless a smooth enough switching function $\chi(\tau)$ with a finite switch-on duration is imposed. The imposition of $\chi(\tau)$ can be viewed as a prescription of regularization to make sense of the perturbation method.

The detector moving along a given trajectory $x(\tau)$ is said to be in equilibrium with the field $\phi$, if
\begin{equation}\label{equilibrium}
D^+(\tau,\tau')\equiv D^+(x(\tau),x(\tau')) = D^+(\Delta\tau), \quad \Delta\tau:=\tau-\tau',
\end{equation}
which depends only on $\Delta\tau$.
In this case, even though \eqref{response function} becomes infinite without the inclusion of $\chi(\tau)$, simply by setting $\chi(\tau)=1$, the (infinite) total transition probability divided by the (infinite) total proper time still sensibly yields a finite \emph{equilibrium transition rate} (i.e., probability per unit proper time) given as
\begin{equation}\label{transition rate}
R=
\kappa^2\abs{\bra{E}m(0)\ket{E_0}}^2
\dot{F}(\Delta E),
\end{equation}
where
\begin{equation}\label{F dot}
\dot{F}(\Delta E) := \int_{-\infty}^\infty d(\Delta\tau)
e^{-i\Delta E\Delta\tau} D^+(\Delta\tau).
\end{equation}
However, the burden of regularization is now carried over to the Wightman function $D^+(x,x')$, which will require some proper regularization procedure, such as the standard $i\epsilon$-regularization used in the case of a detector moving in Minkowski spacetime.\footnote{The standard $i\epsilon$-regularization can be replaced with different regularization procedures, e.g., by imposing a switching function $\chi(\tau)$ or introducing a spatial profile of the Unruh-DeWitt detector. See \cite{Louko:2006zv,Satz:2006kb,Louko:2007mu} for more discussions on the issue of regularization.}
We will make more comments on this point when we encounter the issue of regularization for the Wightman function in \secref{sec:Wightman function}.

On the other hand, if the detector is not in equilibrium with $\phi$ (i.e., $D^+(\tau,\tau')$ depends on both $\tau$ and $\tau'$ for the given trajectory), we can no longer make sense of the notion of equilibrium transition rate but can only refer to the \emph{total} transition probability, which now depends on the exact form of $\chi(\tau)$. Provided that $\chi(\tau)$ is smooth enough and its switch-on duration $\Delta$ is short enough (so that the first-order perturbation is viable), \eqref{transition probability} with \eqref{response function} is well defined and yields a finite total transition probability. By taking the time derivative of the total transition probability, we can still define the \emph{instantaneous transition rate} observed at a particular instant. We refer readers to \cite{Chiou:2016exd,Louko:2006zv,Satz:2006kb,Louko:2007mu} for more discussions on the non-equilibrium case, as we will only focus on the equilibrium case in this paper.

Both the equilibrium transition rate and the instantaneous transition rate in principle can be experimentally measured by deploying a large ensemble of identical Unruh-DeWitt detectors (with the same coupling constant and the same switching function, moving in the same trajectory). Measuring the ratio of the population of the detectors in the ensemble staying in the initial state $\ket{E_0}$ to that of the detectors excited or de-excited to the other state $\ket{E}$, one can deduce the transition rate. (See Appendix A of \cite{Chiou:2016exd} for more details.)

In some situations, the transition process \eqref{transition} and its reverse process $\ket{\Psi,E}\rightarrow\ket{0,E_0}$ can reach \emph{detailed balance}. If the detailed balance is established, the principle of detailed balance dictates that the transition rate $\dot{P}$ of \eqref{transition} and the transition rate $\dot{P}_r$ of its reverse process are both independent of $\tau$ and satisfy
\begin{equation}\label{detailed balance}
\frac{\dot{P}}{\dot{P}_r}=e^{-\beta\Delta E},
\end{equation}
where $1/\beta\equiv k_\mathrm{B}T$ is to be interpreted as the corresponding temperature. The ratio $\dot{P}/\dot{P}_r$ and therefore the temperature of detailed balance in principle can be measured again in terms of a large ensemble of identical detectors (but with a different measuring operation performed upon the ensemble).
Because the amplitudes of the transition process \eqref{transition} and its reverse process are complex conjugate to each other as a consequence of unitarity, the temperature of detailed balance is independent of the explicit design of the detector, as we can see in the examples below.
It should also be remarked that, whereas the condition that the trajectory is in equilibrium with the background field is necessary for detailed balance, it is unclear whether the condition is also sufficient. (See Appendix A of \cite{Chiou:2016exd} and Sec.\ III.A.4 of \cite{Crispino:2007eb}for more discussions about detailed balance.)

In the celebrated example of the Unruh-DeWitt detector moving with a constant acceleration in the Minkowski spacetime, the detailed balance relation is satisfied and the corresponding temperature is given by
\begin{equation}\label{Unruh temperature}
T=\frac{\abs{\text{acceleration}}}{2\pi k_\mathrm{B}},
\end{equation}
which is called the \emph{Unruh temperature}.
This is a consequence of the fact that the Minkowski vacuum is a thermal state of the right (left) Rindler modes (which are the modes of particles seen by the constant-accelerating observer) if the left (right) Rindler modes (which are the modes beyond the apparent event horizon of the constant-accelerating observer) are ignored. More precisely, tracing out the left (right) Rindler modes upon the Minkowski vacuum state gives rise to a density matrix for the many-particle system of the right (left) Rindler modes at the temperature \eqref{Unruh temperature}. (See Secs. III.A.2 and III.A.4 of \cite{Crispino:2007eb} for more details.)
From the perspective of the right (left) Rindler observer, we have
\begin{equation}\label{detailed balance 2}
\frac{\dot{P}}{\dot{P}_r} =\frac{\abs{\mathcal{A}}^2 n(\Delta E)}{\abs{\mathcal{A}_r}^2\left(1+n(\Delta E)\right)}
=\frac{n(\Delta E)}{1+n(\Delta E)}
=e^{-\beta\Delta E},
\end{equation}
where $\mathcal{A}$ and $\mathcal{A}_r$ are the amplitudes measured by the right (left) Rindler observer for the process \eqref{transition} and its reverse process, respectively, which are complex conjugate to each other, and
\begin{equation}
n(\omega) = \frac{1}{e^{\beta\omega}-1}
\end{equation}
is the Rindler particle number density for the thermal state at the Unruh temperature \eqref{Unruh temperature}. The factor $n(\Delta E)$ in the numerator in \eqref{detailed balance 2} is associated with the induced absorption of a Rindler particle from the thermal bath, and the factor $1+n(\Delta E)$ in the denominator is associated with the spontaneous and induced emissions of a Rindler particle to the thermal bath.\footnote{The amplitude of the transition of \eqref{transition} as $\ket{E_0}\rightarrow\ket{E}$ accompanied by the emission of a Minkowski-mode particle into the Minkowski vacuum can be reproduced from the Rindler observer's perspective as accompanied by the absorption of a Rindler-mode particle from the thermal bath (see Sec.~III.A.2 of \cite{Crispino:2007eb} for more details).}

For the case of an Unruh-DeWitt detector moving with a constant velocity in the Minkowski spacetime, from the perspective of a non-moving observer, one can easily compute
\begin{equation}\label{detailed balance 3}
\frac{\dot{P}}{\dot{P}_r} =\frac{\abs{\mathcal{A}}^2 \left(1+n(\Delta E)\right)}{\abs{\mathcal{A}_r}^2n(\Delta E)},
\end{equation}
where $\mathcal{A}$ and $\mathcal{A}_r$ are the amplitudes measured by the non-moving observer, which are complex conjugate to each other, and where
\begin{equation}
n(\omega_{\vec{k}}) :=  \bra{0} a^\dagger_{\vec{k}} a_{\vec{k}} \ket{0} = 0
\end{equation}
is the particle number density for the Minkowski vacuum state $\ket{0}$. The factor $1+n(\Delta E)$ is associated with the spontaneous and induced emissions of a particle to $\ket{0}$, and the factor $n(\Delta E)$ is associated with the induced absorption of a particle from $\ket{0}$. For $\Delta E<0$, it turns out that $\abs{\mathcal{A}}^2=\abs{\mathcal{A}_r}^2\neq0$, and consequently \eqref{detailed balance 3} yields $\dot{P}/\dot{P}_r=\infty$. Therefore, the detailed balance is satisfied in the trivial way corresponding to the zero temperature $T=0$ (i.e., $\beta=\infty)$. For $\Delta E>0$, it turns out that $\abs{\mathcal{A}}^2=\abs{\mathcal{A}_r}^2=0$, and the temperature is ill-defined.

In this paper, we study the Unruh-DeWitt detector moving along a free-falling trajectory or along a constant-accelerating trajectory in a monochromatic gravitational wave background, instead of the Minkowski background. As we will see, in various settings in the long-wavelength or short-wavelength limit of the gravitational wavelength, a free-falling or constant-accelerating detector is in equilibrium with $\phi$. The equilibrium transition rate depends on the amplitude of the gravitational wave, which can be viewed as a measurable effect of gravitational waves acting on a quantum system.
However, we do not consider detailed balance and the corresponding temperature. It is unclear whether detailed balance can be established in the presence of a gravitational wave. Even if detailed balance is established in certain settings, it is difficult to obtain the particle number density $n(\omega)$ in the gravitational wave background that can be used to compute the ratio $\dot{P}/\dot{P}_r$. We leave the issue of detailed balance for future research.

Finally, we remark that the underlying mechanism of the transition \eqref{transition} is that the Unruh-DeWitt detector is coupled to quantum fluctuations of $\phi$ in the vacuum $\ket{0}$, akin to spontaneous emission of an atom or a molecule as a consequence of being coupled to quantum fluctuations of the electromagnetic field. From the experimental perspective, it seems more realistic to model the detector as coupled to the electromagnetic field, rather than a scalar field $\phi$, as Nature evidently has quantum fluctuations in the vacuum of the electromagnetic field, but may not of a scalar field.
Indeed, generalized Unruh-DeWitt models coupled with different kinds of quantum fields have been formulated, including the electromagnetic field \cite{Boyer:1980wu,Boyer:1984yqq} and the Dirac field \cite{Iyer:1980yc}. However, in this paper, we adhere to the original Unruh-DeWitt model with a scalar field, as it is theoretically the simplest and its simplicity enables us to obtain the transition rate in a closed form that is easier to analyze.
The transition rate of the Unruh-DeWitt detector coupled to a non-scalar field in general depends not only on the detector's trajectory but also its orientation, as it can be sensitive to polarization of the filed. On the other hand, when a large ensemble of detectors is used to measure the transition rate, the ensemble as a whole may become insensitive to the field's degrees of polarization, if each detector in the ensemble is randomly oriented. The transition rate measured by the randomly-oriented ensemble as a whole can be represented by the simple model coupled to a scalar field (up to some detailed dependence on the explicit form of coupling). It is in this sense that the simple model with a scalar field is still relevant to realistic concern.

\section{Scalar field in a gravitational wave background}\label{sec:scalar field}
In this section, we solve the equation of motion of a real scalar field, i.e.\ the Klein-Gordon equation, in a gravitational wave background, which is otherwise a flat spacetime in the absence of gravitational waves. The gravitational waves are assumed to be weak enough so that the linearized theory, which neglects nonlinear gravitational wave effects, is adequate. In the linearized theory, any arbitrary gravitational wave can be decomposed into a linear superposition of plane waves. For simplicity, we only consider a monochromatic plane wave. To make the calculation simpler, we work in the transverse-traceless (TT) gauge.\footnote{For the issue that the TT gauge is always possible for any arbitrary gravitational wave, see Section 35.4 of \cite{Misner:1974qy} for more details.}

The action of a scalar field $\phi(x)$ in a curved spacetime is given by
\begin{equation}
S = \int \mathcal{L}[\phi(x)]\, d^4x
\end{equation}
with the Lagrangian density
\begin{equation}\label{L in curved spacetime}
\mathcal{L} =
\frac{1}{2}\sqrt{-g}
\left(- g^{\mu\nu} \nabla_\mu\phi \nabla_\nu\phi
        - m^2\phi^2
        - \xi R \phi^2
\right),
\end{equation}
where $m$ is the mass of the scalar particle, $R$ is the Ricci scalar, and $\xi$ is the coupling constant for the interaction between $\phi$ and $R$ (see e.g.\ \cite{Birrell:1982ix} for more details).
In order to obtain the Wightman function in a closed form, we consider the simplest case that $\phi$ is massless and does not couple with the curvature, i.e. $m=0$ and $\xi=0$. Variation with respect to $\phi$, i.e.\ $\delta S/\delta\phi=0$, then leads to the massless Klein-Gordon equation in curved spacetime,
\begin{equation}
\square \phi=0,
\end{equation}
where
\begin{eqnarray}\label{KG}
\square \phi
&:=& g^{\mu\nu}\nabla_\mu\nabla_\nu\phi
\equiv \frac{1}{\sqrt{-g}}\, \partial_\mu \left(\sqrt{-g}\, g^{\mu\nu} \partial_\nu \phi \right)\nonumber \\
&=&
g^{\mu\nu}\partial_\mu\partial_\nu\phi
+\partial_\mu g^{\mu\nu} \partial_\nu \phi
+\frac{1}{2}g^{\mu\nu} g^{\alpha\beta}\partial_\mu g_{\alpha\beta}\partial_\nu \phi.
\end{eqnarray}

The metric of the spacetime with a gravitational plane wave is given by
\begin{subequations}
\begin{eqnarray}
g_{\alpha\beta}(x)&=&\eta_{\alpha\beta}+h_{\alpha\beta}(x),\\
h_{\alpha\beta}(x)&=&A_{\alpha\beta}\,e^{-ik_\mu x^\mu}.
\end{eqnarray}
\end{subequations}
It follows that
\begin{subequations}
\begin{eqnarray}
g^{\alpha\beta} \partial_\mu g_{\alpha\beta}
&=& g^{\alpha\beta} \partial_\mu   h_{\alpha\beta}=  k_\mu g^{\alpha\beta}h_{\alpha\beta}= k_\mu  h_{\alpha\beta}h^{\alpha\beta}\approx O(h^2),\\
\partial_\mu g^{\mu\nu}&=&\partial_\mu h^{\mu\nu}= k_\mu h^{\mu\nu}  =0,
\end{eqnarray}
\end{subequations}
where we have applied the TT gauge to have $k_\mu h^{\mu\nu}=0$.
Consequently, up to the first order of $h$, the Klein-Gordon equation reads as
\begin{equation}
  g^{\mu\nu}\partial_\mu\partial_\nu\phi=0. 
\end{equation}

Given a gravitational plane wave prorogating along the $z$ direction, in the TT gauge, $h_{\mu\nu}$ takes the form
\begin{subequations}\label{h}
\begin{eqnarray}
h_{\mu\nu}(x)&=&\left(
            \begin{array}{cccc}
              0 & 0 & 0 & 0 \\
              0 & h_{+} & h_{\times} & 0 \\
              0 & h_{\times} & -h_{+} & 0 \\
              0 & 0 & 0 & 0
            \end{array}
           \right),\\
h_{+/\times}(x)&=&A_{+/\times} \cos( kz- \omega t +\theta_{+/\times}),
\end{eqnarray}
\end{subequations}
where $\omega = \abs{k}$, and $h_+$ and $h_\times$ are the two independent modes of polarization with the amplitudes $A_+$ and $A_\times$ and phase shifts $\theta_+$ and $\theta_\times$, respectively.
Note that the metric given in \eqref{h} admits the Killing vector fields: $X=\partial_x$, $Y=\partial_y$, and $V=\partial_v=(\partial_t+\partial_z)/\sqrt{2}$. Accordingly, it is privileged to introduce the \emph{light-front} variables:
\begin{equation}\label{lc}
u=(t-z)/\sqrt{2},\quad v=(t+z)/\sqrt{2}.
\end{equation}
In terms of the coordinates $(u,v,x,y)$, the Klein-Gordon equation in the gravitational wave background reads as
\begin{eqnarray}\label{KG in LC}
&&\left[
    -2   \frac{\partial^2}{\partial u\partial v}
    +    \frac{\partial^2}{\partial x^2}
    +    \frac{\partial^2}{\partial y^2}
    -  A_+\cos(w u+\theta_+)\left(\frac{\partial^2}{\partial x^2}-\frac{\partial^2}{\partial y^2}\right)
\right.\nonumber\\
&&
\left.
    \qquad\qquad\qquad\qquad\qquad
    \mbox{} - 2A_\times\cos(w u+\theta_\times)   \frac{\partial^2}{\partial x\partial y}
    \right] \phi
=0,
\end{eqnarray}
where we define the shorthand notation $w$ as
\begin{equation}\label{lc}
w:=\sqrt{2}\,\omega.
\end{equation}

As \eqref{KG in LC} is invariant under the translations along the $x$, $y$, and $v$ coordinates in accordance with the Killing vectors, we make the ansatz that the solution of $\phi$ takes the form
\begin{equation}\label{ansatz}
  \phi(u,v,x,y) = e^{i(k_xx+k_yy -\omega v)} \chi(u),
\end{equation}
where $\chi(u)$ is a function of $u$ to be determined.
Substituting \eqref{ansatz} into \eqref{KG in LC}, we have
\begin{equation}
  2i\,\omega \chi'(u)= \left[k_x^2 +k_y^2-A_+\cos(w u+\theta_+)(k_x^2- k_y^2) -2 A_\times\cos(w u+\theta_\times) k_x k_y\right] \chi(u).
\end{equation}
Integrating this equation then yields
\begin{equation}\label{chi}
\chi(u)
\propto
e^{-i k_u u } \,
e^{i k_u g_c(k_x,k_y)\frac{\sin{w u}}{w}}
e^{i k_u g_s(k_x,k_y)\frac{\cos{w u}}{w}},
\end{equation}
where $k_u:=(k_x^2+k_y^2)/{2\omega}$, or, equivalently, the mode frequency as a function of $k_u$, $k_x$, and $k_y$ is given by
\begin{equation}\label{omega k}
\omega\equiv
\omega_{k_u,k_x,k_y}
:=\frac{k_x^2+k_y^2}{2k_u},
\end{equation}
and
\begin{subequations}
\begin{eqnarray}
g_c(k_x,k_y)
&:=& \frac{1}{(k_x^2+k_y^2)}
    \left[ A_+  (k_x^2-k_y^2)\cos\theta_+
        + 2A_\times k_xk_y \cos\theta_\times\right] ,  \\
g_s(k_x,k_y)
&:=& \frac{1}{(k_x^2+k_y^2)}
    \left[ A_+  (k_x^2-k_y^2)\sin\theta_+
        + 2A_\times k_xk_y \sin\theta_\times\right].
\end{eqnarray}
\end{subequations}
For a given gravitational plane wave parametrized by $A_{+,\times}$, $\theta_{+,\times}$, and $w\equiv\sqrt{2}\,\omega$, we have obtained the eigenmode solutions of $\phi(u,v,x,y)$ parametrized by $k_x$, $k_y$, and $k_u$. Based on these eigenmodes, we can perform the field quantization in the next section.

\section{Light-front quantization}\label{sec:LF quantization}
The ordinary equal-time quantization scheme in curved spacetime requires a timelike Killing vector field to make sense of the notion of time \cite{Birrell:1982ix}. As the metrics given by \eqref{h} exhibits two spacelike and one lightlike Killing vector fields, but no timelike one, the ordinary scheme cannot apply. Instead, we adopt the \emph{light-front quantization} formalism, which is primarily used in the study of deep inelastic scattering in quantum chromodynamics (QCD) (see \cite{Burkardt:1995ct} for a review).
Under the light-front quantization scheme, the light-front direction in accordance with the lightlike Killing vector is treated as the direction of time. More precisely, we treat the light-front coordinate $v$ as the evolution parameter, and correspondingly define the frequency modes as eigenmodes of the Lie derivative via
\begin{equation}
  \mathcal{L}_{V} \phi = -i \omega_k\phi, \quad \text{with}\,  V=\partial_v,
\end{equation}
where $\omega_k$ is the frequency.

The Lagrangian density \eqref{L in curved spacetime} in the gravitational wave background given by \eqref{h} takes the form $\mathcal{L}=\sqrt{-g} \, (\partial_u \phi \partial_v \phi +\dots)$ in the coordinates $(u,v,x,y)$, where the part of ``$\dots$'' does not involve $\partial_v\phi$.
Consequently, the canonical momentum conjugate to $\phi$ is given by
\begin{equation}\label{pi}
\pi
:=\frac{\partial\mathcal{L}}{\partial(\partial_v\phi)}= \sqrt{-g}\,\partial_u\phi.
\end{equation}
The light-front quantization then demands the commutation relations given at equal light-front time $v$ as
\begin{equation}\label{CR 1}
[\phi(\bm{x},u,v),\pi(\bm{x}',u',v)]
=
\frac{i}{2} \sqrt{-g}\,\delta^2(\bm{x}-\bm{x}')\,\delta(u-u'),
\end{equation}
and
\begin{equation}\label{CR 2}
[\pi(\bm{x},u,v),\pi(\bm{x}',u',v)]=0.
\end{equation}
Note that \eqref{CR 1} implies the nonlocal commutation relation,
\begin{equation}
[\phi(\bm{x},u,v),\phi(\bm{x}',u',v)]
=
-\frac{i}{4} \sqrt{-g}\,\delta^2(\bm{x}-\bm{x}')\,\mathrm{sgn}(u-u'),
\end{equation}
which is a new feature that does not appear in the ordinary equal-time quantization scheme.\footnote{\label{foot:Dirac algorithm}Note that \eqref{pi} does not contain any time derivative $\partial_v$, and thus should be considered as a constraint equation. In other words, the phase space variables $\phi(x)$ and $\pi(x)$ at a given time $v$ are not completely independent of each other. In the presence of constraints, one has to apply the Dirac-Bergmann algorithm to arrive at a consistent Hamiltonian formalism, which then provides a proper starting point for the quantization procedure. Following the Dirac-Bergmann procedure, it is the Dirac bracket, instead of the Poisson bracket, that is to be promoted to the commutator $[\,\cdot,\cdot\,]$ for quantization. The difference between the Dirac bracket and the Poisson bracket gives rise to the extra factor $1/2$ in \eqref{CR 1}. See Appendix of \cite{Burkardt:1995ct} for more details.
In the light-front formalism, one has to address the additional issue arising from \emph{zero modes}, which correspond to the states that are independent of $u$ and have to be treated separately with special care. The resulting modified Dirac–Bergmann procedure could be very complicated, as the main difficulty lies in the fact that the constraint equation for zero modes is generally nonlinear. Fortunately, in our case as well as in many cases of free theories, the zero mode constraint does not get involved with the Hilbert space orthogonal to the zero modes, and thus can be simply projected out before the standard Dirac-Bergmann procedure is applied. For more about the zero-mode problem, see Appendix of \cite{Burkardt:1995ct} and the references therein.}
Here and hereafter, we use boldfaced letters to denote ``transverse'' vectors in shorthand: e.g., $\bm{x}:=(x,y)$, $\bm{k}:=(k_x,k_y)$, and $\bm{k}\cdot\bm{x}:= k_x x + k_yy$.

By virtue of \eqref{ansatz} and \eqref{chi}, the fields $\phi(x)$ can be cast in terms of the mode expansions as
\begin{equation}\label{phi of a and a dag}
\phi (u,v,\bm{x})
=\int \frac{d^3k}{(2\pi)^{3/2}}  \,N_{k_u,\bm{k}}\,
                 \left(   a_{k_u,\bm{k}}   \,f_{k_u,\bm{k}}(\bm{x},u,v)
                        + a^\dag_{k_u,\bm{k}}  \,f^*_{k_u,\bm{k}}(\bm{x},u,v)\right),
\end{equation}
where
\begin{equation}
f_{k_u,\bm{k}} (u,v,\bm{x}):=e^{ i(\bm{k}\cdot \bm{x}-k_uu -\omega_{k_u,\bm{k}}v)} \chi(u)
\end{equation}
with $\chi(u)$ given by \eqref{chi}, and where $N_{k_u,\bm{k}}$ are normalization factors to be determined later.
The conjugate momentum field $\pi$ given by \eqref{pi} then reads as
\begin{eqnarray}
\pi (u,v,\bm{x})
&=&\sqrt{-g} \int \frac{d^3k}{(2\pi)^{3/2}}\, N_{k_u,\bm{k}}\,(-ik_u)\,g_{\bm k}(u)\\ \nonumber
&&\qquad\qquad\qquad \times\left(   a_{k_u,\bm{k}}       \,f_{k_u,\bm{k}}(\bm{x},u,v)
                        - a^\dag_{k_u,\bm{k}}  \,f^*_{k_u,\bm{k}}(\bm{x},u,v)\right),
\end{eqnarray}
where
\begin{equation}
g_{\bm k}(u):=1- g_c(k_x,k_y)\cos(w u)+g_s(k_x,k_y)\sin(w u).
\end{equation}
Note that $k_u\geq0$ according to \eqref{omega k},\footnote{We have adopted the convention that $\omega_{k_u,\bm{k}}\geq0$. That is, positive-frequency modes (i.e., $\propto e^{i\omega_{k_u,\bm{k}}v}$) are associated with creation operators, while negative-frequency (i.e., $\propto e^{-i\omega_{k_u,\bm{k}}v}$) with annihilation operators in \eqref{phi of a and a dag}.} and the notation $\int d^3k$ is a shorthand for $\int_{-\infty}^{\infty} d^2k \int_{0}^{\infty} dk_u \equiv \int_{-\infty}^{\infty} dk_x \int_{-\infty}^{\infty} dk_y \int_{0}^{\infty} dk_u$.

By prescribing the commutation relations for $a_{k_u,\bm{k}}$ and $a^\dag_{k_u,\bm{k}}$ as
\begin{subequations}\label{a and a dag}
\begin{eqnarray}
[a_{k_u,\bm{k}},a^\dag_{k'_u,\bm{k}'}] &=& \delta^2(\bm{k}-\bm{k}')\delta(k_u-k'_u),\\
{[}a_{k_u,\bm{k}},a_{k'_u,\bm{k}'}{]} &=& [a^\dag_{k_u,\bm{k}},a^\dag_{k'_u,\bm{k}'}] = 0,
\end{eqnarray}
\end{subequations}
and the normalization factor as
\begin{equation}
N_{k_u,\bm{k}}=\frac{1}{\sqrt{2k_u}},
\end{equation}
the commutation relations \eqref{CR 1} and \eqref{CR 2} can be realized.
To show this, we first calculate
\begin{eqnarray}
&&[\phi(\bm{x},u,v),\pi(\bm{x}',u',v)]\nonumber\\
&=&\sqrt{-g}
\int \frac{d^3k}{(2\pi)^{3/2}}  \int \frac{d^3k'}{(2\pi)^{3/2}}  \,\frac{i g_{\bm k}(u')k_u'}{\sqrt{4k_uk'_u}} \,
        \bigg\{ [a_{k_u,\bm{k}}, a^\dag_{k'_u,\bm{k}'} ]\,f_{k_u,\bm{k}}(\bm{x},u,v) f^*_{k'_u,\bm{k}'}(\bm{x}',u',v) \nonumber\\
&&\mbox{       } +
        [a_{\bm{k}',k_u},a^\dag_{\bm{k},k_u } ]\,f_{k'_u,\bm{k}'}(\bm{x'},u',v)  f^*_{k_u,\bm{k}}(\bm{x},u,v)                 \bigg\}\nonumber\\
&=&
    \frac{i}{4}\sqrt{-g}\int_{-\infty}^{\infty} \frac{d^2k}{(2\pi)^2} \,g_{\bm k}(u')
    \left(         e^{ i\bm{k}\cdot (\bm{x}-\bm{x}')}+ e^{-i\bm{k}\cdot (\bm{x}-\bm{x}')}        \right)
    \int_{-\infty}^{\infty} \frac{dk_u}{2\pi}  e^{-ik_u\lambda(u)}
\nonumber\\
&=&
    \frac{i}{4}\sqrt{-g}\, \int_{-\infty}^{\infty} \frac{d^2k}{(2\pi)^2} \,\delta(\lambda(u))\, g_{\bm k}(u')
    \left(        e^{ i\bm{k}\cdot (\bm{x}-\bm{x}')}+ e^{-i\bm{k}\cdot (\bm{x}-\bm{x}')}        \right),
\end{eqnarray}
where
\begin{equation}
\lambda(u) := u-u'+ w^{-1}\left[ g_c(k_x,k_y)(\sin{w u}-\sin{w u'})+ g_s(k_x,k_y)(\cos{w u}-\cos{w u'})\right].
\end{equation}
Since $\lambda(u)$ has a single root at $u=u'$, the identity
\begin{equation}
 \delta(\lambda(u)) = \frac{\delta(u-u')}{\abs{\lambda'(u')}} =\frac{\delta(u-u')}{g_{\bm k}(u')}
\end{equation}
can be used to obtain
\begin{eqnarray}
[\phi(\bm{x},u,v),\pi(\bm{x}',u',v)]
&=& \frac{i}{4} \sqrt{-g}\,\delta(u-u') \int_{-\infty}^{\infty} \frac{d^2k}{(2\pi)^2} \,
    \left( e^{ i\bm{k}\cdot (\bm{x}-\bm{x}')}+ e^{-i\bm{k}\cdot (\bm{x}-\bm{x}')}        \right),\nonumber\\
&=& \frac{i}{2} \sqrt{-g}\,\delta^2(\bm{x}-\bm{x}')\,\delta(u-u'),
\end{eqnarray}
in agreement with \eqref{CR 1}.
Meanwhile, it is can be readily shown that \eqref{a and a dag} leads to \eqref{CR 2}.

In summary, the mode expansion of $\phi(x)\equiv\phi(u,v,\bm{x})$ in terms of creation and annihilation operators is given by
\begin{equation}\label{phi}
\phi(x)
=\int \frac{dk^3}{(2\pi)^{3/2}}  \,\frac{1}{\sqrt{2k_u}}\,
                 \left(   a_{k_u,\bm{k}}       \,f_{k_u,\bm{k}}(u,v,\bm{x})
                        + a^\dag_{k_u,\bm{k}}  \,f^*_{k_u,\bm{k}}(u,v,\bm{x})\right).
\end{equation}
The Hilbert space is spanned by the Fock states in the form
\begin{eqnarray}
&&\ket{{}^1n_{k_{u1},\bm{k}_1},{}^2n_{k_{u2},\bm{k}_2},\dots,{}^jn_{k_{uj},\bm{k}_j}} \nonumber\\
&:=&({}^1n!\,{}^2n!\dots{}^jn!)^{-1/2}
(a^\dag_{k_{u1},\bm{k}_1})^{{}^1n}(a^\dag_{k_{u2},\bm{k}_2})^{{}^2n}\dots(a^\dag_{k_{uj},\bm{k}_j})^{{}^jn}
\ket{0},
\end{eqnarray}
which is a many-particle state with ${}^1n$ particles in the mode $(k_{u1},\bm{k}_1)$, ${}^2n$ particles in the mode $(k_{u2},\bm{k}_2)$, and so on. The no-particle state $\ket{0}$ is the vacuum, which is annihilated by all annihilation operators, i.e.,
\begin{equation}
  a_{k_u,\bm{k}}\ket{0}=0,\quad  \text{for all}\ (k_u,\bm{k}).
\end{equation}

\section{The Wightman function}\label{sec:Wightman function}
As we have successfully quantized the scalar field $\phi(x)$ in a monochromatic gravitational wave background, we are now ready to calculate the corresponding Wightman function.
According to \eqref{phi}, the Wightman function $D^+(x,x')$ takes the form
\begin{eqnarray}\label{Wightman}
D^+(x,x')
&:=& \bra{0}\phi(x)\phi(x')\ket{0} \nonumber\\
&=&
\int \frac{dk^3}{(2\pi)^{3/2}}
\int \frac{dk'^3}{(2\pi)^{3/2}} \,
\frac{1}{\sqrt{4k_u k'_u}}  \,
\bra{0} a_{k_u,\bm{k}} a^\dag_{k'_u,\bm{k}'} \ket{0}\,
f_{k_u,\bm{k}}f^*_{k_u,\bm{k}}
\nonumber\\
&=&
\int \frac{dk^3}{(2\pi)^3}\frac{1}{2k_u}
e^{ i\bm{k}\cdot (\bm{x}-\bm{x}')} e^{-ik_u(u-u')} e^{-i\omega_{k_u,\bm{k}}(v-v')}\nonumber\\
&&\mbox{}\times
e^{ik_u g_c(\bm{k})(\sin{w u}-\sin{w u'})/w}\,
e^{ik_u g_s(\bm{k})(\cos{w u}-\cos{w u'})/w}.
\end{eqnarray}
This expression is complicated and does not have a closed form. Fortunately, it can be greatly simplified if we consider the two limiting situations: the \emph{long-wavelength limit} and the \emph{short-wavelength limit}.

The Unruh-DeWitt detector naturally provides two characteristic timescales. The first is ${\sim}1/\Delta E$, which characterizes the detector's response time for the two-level transition. The second is ${\sim}\Delta$, which characterizes the detector's switch-on period as shown in \figref{fig:switching function}.
As will be seen shortly, we will apply a particular form of $i\epsilon$-regularization to compute the Wightman function instead of specifying the switching function $\chi(\tau)$ with a finite switch-on period $\Delta$. In reality, however, a detector is always switched on only for a finite period $\Delta$, which lays down a timescale to be compared with the period $1/\omega$ of the gravitational wave. Therefore, we should keep in mind that the resulting transition rate of the Unruh-DeWitt detector computed from the $i\epsilon$-regularized Wightman function is legitimate only if the measurement performed at time $\tau$ is well within the switch-on period.\footnote{Accordingly, the condition $1/\Delta E\ll\Delta$ must be satisfied in order to yield a sensible result in agreement with the transition rate computed from the regularized Wightman function.}

\begin{figure}
\centering
  \begin{minipage}[b]{0.4\textwidth}
    \includegraphics[width=\textwidth]{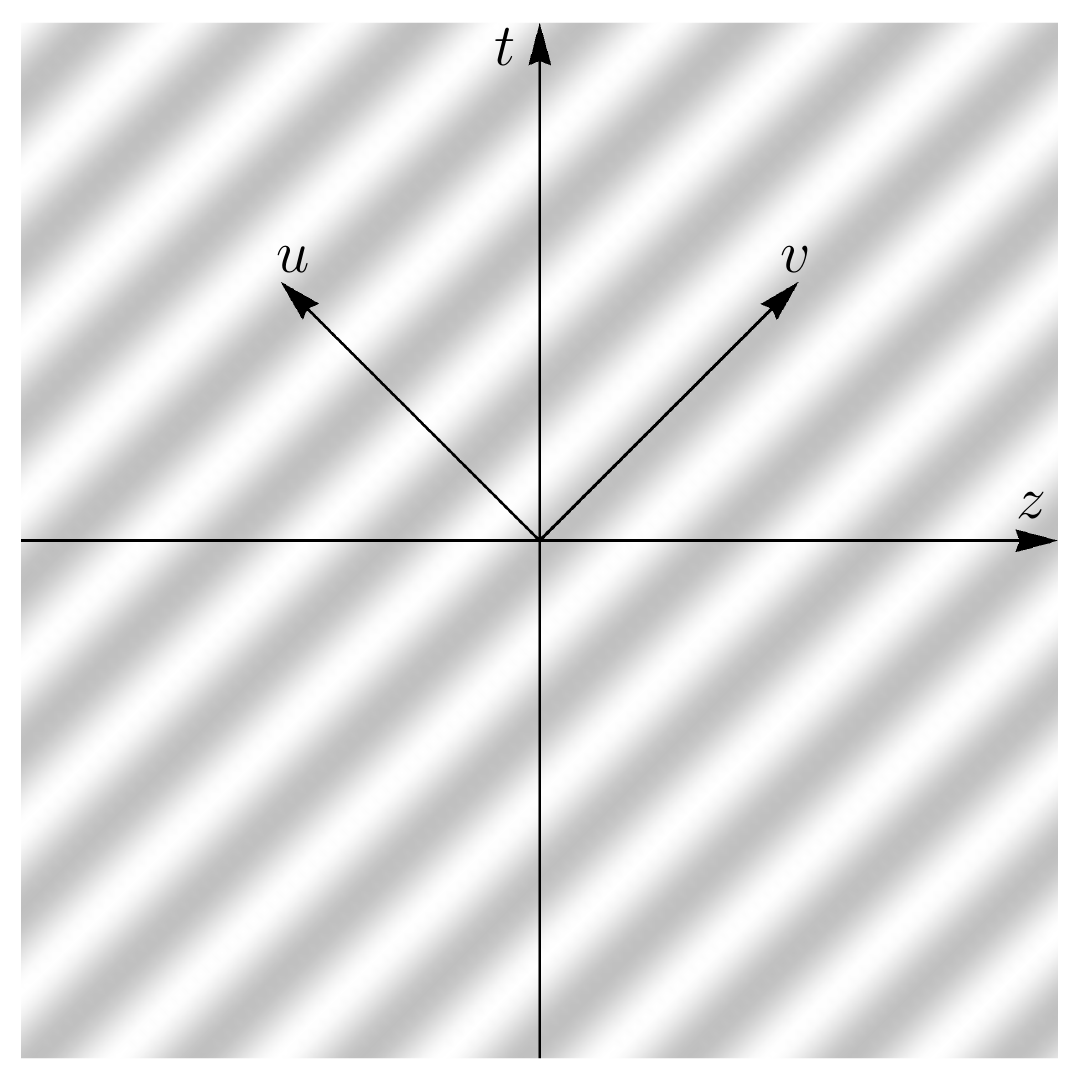}
  \end{minipage}
  \hspace{0.7cm} 
  \begin{minipage}[b]{0.4\textwidth}
    \includegraphics[width=\textwidth]{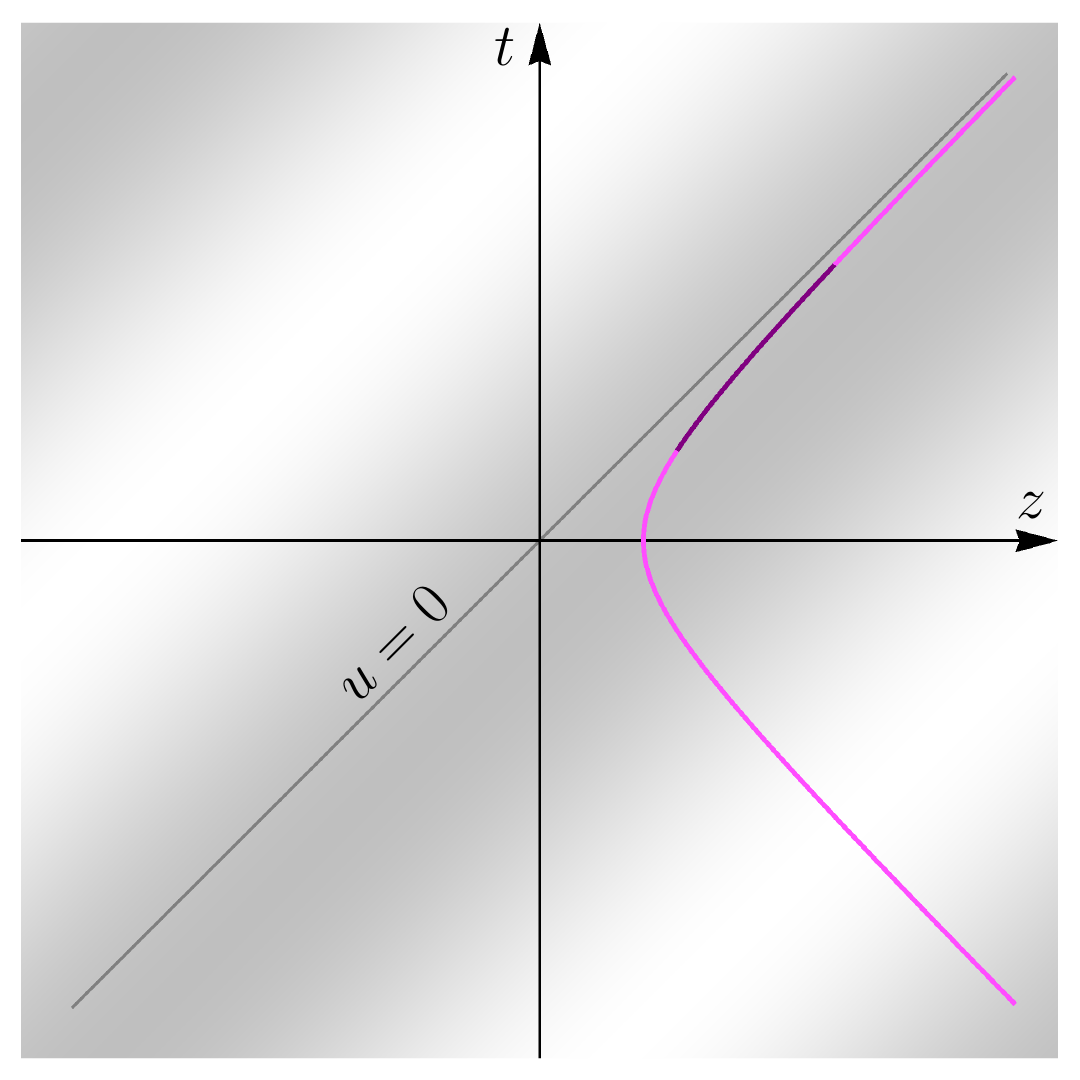}
  \end{minipage}
\caption{[\textit{Left}] Modulation of the phase of a gravitational plane wave propagating in the $z$ direction. The axes of $(t,z)$ and $(u,v)$ are both shown for reference. [\textit{Right}] A case of a very long wavelength is drawn to illustrate the long-wavelength condition \eqref{long-wavelength condition}. A world line given by \eqref{x a} (with $t_0=0$) is depicted as an example of the detector's trajectory, and the segment within the switch-on period, $\tau_0\lesssim\tau\lesssim\tau_0+\Delta$, is highlighted in a deeper shade (with $\tau_0$ chosen to be a certain positive value). The wavelength is so large that the phase of the gravitational wave is almost the same over the whole switch-on segment. Furthermore, if $\tau_0$ is chosen to be positive and very large, the switch-on segment will asymptote to the line of $u=0$, and the corresponding velocity will asymptote to $v_z\approx1$. In this asymptotic situation, the long-wavelength condition is satisfied even if $1/\omega$ is small.}
\label{fig:long-wavelength}
\end{figure}

The background gravitational wave is said to be in the long-wavelength limit, if the wavelength of the gravitational wave is so large that, within the whole switch-on period, the detector does not see any gravitational wave modulation, but effectively only sees a persisting gravitational wave amplitude. See \figref{fig:long-wavelength} for illustration.
More precisely, during the switch-on period, if the detector moves from $x^\mu(\tau)=(t,\vec{x})\equiv(t,x,y,z)$ to $x^\mu(\tau+\Delta)=(t',\vec{x}')\equiv(t',x',y',z')$, we have
\begin{equation}
t' \approx t + \frac{1}{1-v^2}\Delta,
\qquad
z' \approx z + \frac{v_z}{1-v^2}\Delta,
\end{equation}
where $\vec{v}=(v_x,v_y,v_z)$ is the detector's moving velocity (averaged over the switch-on period). The phase difference of the gravitational wave experienced by the detector during this period is given by
\begin{equation}
\delta\phi = (kz'-\omega t') - (kz -\omega t)
\approx \frac{\omega\Delta}{\sqrt{1-v^2}}(v_z-1),
\end{equation}
for a gravitational plane wave propagating in the $z$ direction ($k=\omega>0$). The precise condition for the long-wavelength limit is $\abs{\delta\phi}\ll1$, or equivalently,
\begin{equation}\label{long-wavelength condition}
\frac{1}{\Delta E} \ll \Delta \ll \frac{\sqrt{1-v^2}}{\abs{v_z-1}}\,\frac{1}{\omega}.
\end{equation}
It should be noted that, in the case of a detector moving in the $z$ direction at an extremely fast speed close to the speed of light, i.e., $v_z\approx1$, the condition \eqref{long-wavelength condition} is always satisfied even if the gravitational wavelength $1/\omega$ is small. The asymptotic behavior of the constant-accelerating trajectory given by \eqref{x a} is a typical example (see \figref{fig:long-wavelength}). At the opposite extreme, if the detector moves in the negative $z$ direction at a speed close to the speed of light, i.e., $v_z\approx-1$, the condition \eqref{long-wavelength condition} cannot be satisfied, no matter how long $1/\omega$ is.

If we neglect any corrections equal to or higher than the order of $O(\omega\Delta)$, we can simply take the formal limit $\omega\rightarrow0$ for the result of the long-wavelength limit.
In this formal limit, wherever the portion of the trajectory within the switch-on period is located in the spacetime, the phase of the gravitational wave upon this portion is to be treated as the same as that upon the hypersurface $u=0$. (See the right panel of \figref{fig:long-wavelength}, imaging that the wavelength becomes infinity.)
Therefore, the persisting amplitude the detector experiences during the switch-on period is given by $\left.h_{+/\times}(x)\right|_{u=0}\equiv\mathcal{A}_{+/\times}$ for $+$ and $\times$ modes, respectively, which is defined as
\begin{equation}\label{cal A}
 \mathcal{A}_+:= A_+\cos \theta_+,\quad \mathcal{A}_\times:= A_\times\cos\theta_\times.
\end{equation}
In a real \emph{physical} setting (contrary to the \emph{formal} limit $\omega\rightarrow0$), if the long-wavelength condition \eqref{long-wavelength condition} satisfied, $\mathcal{A}_+$ and $\mathcal{A}_\times$ used in the formal limit are to be understood as representing the persisting amplitudes experienced by the detector during the switch-on period.

On the other hand, the background gravitational wave is said to be in the short-wavelength limit, if the condition
\begin{equation}\label{short-wavelength condition}
\frac{\sqrt{1-v^2}}{\abs{v_z-1}}\,\frac{1}{\omega} \ll \frac{1}{\Delta E} \ll \Delta
\end{equation}
is satisfied. That is, the frequency of gravitational-wave modulation experienced by the moving detector is much higher than the frequency of the two-level energy difference. If we neglect any corrections equal to or higher than the order of $O(\Delta E/\omega)$, we can simply take the formal limit $\omega\rightarrow\infty$ for the result of the short-wavelength limit.

Although we cannot obtain a closed-form expression for the response of the Unruh-DeWitt detector in the case of arbitrary wavelengths, we can still learn a great deal from the two opposite limits.

\subsection{Long-wavelength limit}
We first consider the long-wavelength limit conditioned by \eqref{long-wavelength condition}. Neglecting any corrections in or higher than the order of $O(\omega\Delta)$, we take the formal limit $w\equiv\sqrt{2}\,\omega\rightarrow 0$ upon \eqref{Wightman}:
\begin{eqnarray}\label{L limit}
D^+_{\mathrm{lw}}(x-x')
&\equiv& \lim_{w\rightarrow0}D^+(x,x') \nonumber\\
&=&
\int \frac{d^3k}{(2\pi)^3}\frac{1}{2k_u}
e^{ i\bm{k}\cdot (\bm{x}-\bm{x}')}e^{-ik_u(1-g_c(k_x,k_y))(u-u')}e^{ -i\omega_{k_u,\bm{k}} (v-v')}.
\end{eqnarray}
By performing the change of variables,
\begin{subequations}
\begin{eqnarray}
k_u' &\equiv& k_u'(k_u,k_x,k_y)
=k_u (1-g_c(k_x,k_y)),\\
\omega_{k_u',\bm{k}}
&\equiv& \omega(k_u',k_x,k_y)
= \omega_{k_u,\bm{k}}
= \frac{k_x^2+k_y^2}{2k'_u}\left[1-g_c(k_x,k_y)\right],
\end{eqnarray}
\end{subequations}
\eqref{L limit} is simplified into the form
\begin{equation}\label{Wightman lw}
D^+_{\mathrm{lw}}(x-x')
=\int \frac{d^2k\,dk_u'}{(2\pi)^3}\frac{1}{2k_u'}
e^{ i\bm{k}\cdot (\bm{x}-\bm{x}')}e^{-ik_u'(u-u')}e^{ -i\omega_{k_u',\bm{k}} (v-v')},
\end{equation}
where we have used $dk_u=(1-g_c(k_x,k_y))^{-1}dk_u'$.
The Wightman function is in fact not a genuine function but a distribution. When substituted into \eqref{response function}, it yields an unambiguous result for the response function $F(\Delta E)$ as long as the switching function $\chi(\tau)$ is smooth enough and of compact support. Without specifying $\chi(\tau)$, however, the expression \eqref{Wightman lw} by itself is ambiguous and requires a proper regularization procedure to yield a sensible result in agreement with the condition of causality.\footnote{Recall the comments after \eqref{F dot}.} Here, we prescribe the particular form of $i\epsilon$-regularization (with an infinitesimal parameter $\epsilon>0$) as follows:
\begin{equation}\label{Wightman lw regularized}
D^+_{\mathrm{lw}}(x-x')
=\int \frac{d^2k\,dk_u'}{(2\pi)^3}\frac{1}{2k_u'}
e^{ i\bm{k}\cdot (\bm{x}-\bm{x}')}e^{-ik_u'(u-u'-i\epsilon)}e^{ -i\omega_{k_u',\bm{k}} (v-v'-i\epsilon)}.
\end{equation}
This $i\epsilon$-regularization conforms with the condition of causality, as we will see shortly that it reduces to the standard $i\epsilon$-regularization when the gravitational wave amplitude is turned off. This particular regularization can also be understood as providing a large-value cutoff for both $\omega_{k_\mu,\bm{k}}$ and $k_u$.

Applying the Gaussian integral formula
\begin{equation}
\int_{-\infty}^{\infty} dx e^{-ax^2+bx+c}=\sqrt{\frac{\pi}{a}}\, e^{\frac{b^2}{4a}+c}
\end{equation}
to the integration over $k_x$ and $k_y$, we have
\begin{eqnarray}\label{Iku}
\mathcal{I}(k'_u)
&:=&\int \frac{dk_x}{2\pi} \int\frac{dk_y}{2\pi}\,
e^{ i k_x \Delta x}\, e^{ i k_y \Delta y}\, e^{ -i\omega_{k'} (v-v'-i\epsilon)} \nonumber\\
&=&
\frac{k'_u}{2\pi i}
\frac{1}{\sqrt{1-\mathcal{A}^2}}
\frac{1}{(v-v'-i\epsilon)}\,
e^{i k'_u\, R(\Delta x,\Delta y)\,/2(v-v'-i\epsilon)},
\end{eqnarray}
where $\Delta x:=x-x'$, $\Delta y:=y-y'$, and $R(\Delta x,\Delta y)$ is defined as
\begin{equation}
R(\Delta x,\Delta y)
:=
\frac{(1+\mathcal{A}_+) \Delta x^2 + 2 \mathcal{A}_\times\Delta x\Delta y+ (1-\mathcal{A}_+)\Delta y^2}{1-\mathcal{A}^2},
\end{equation}
with $\mathcal{A}_+$ and $\mathcal{A}_\times$ defined in \eqref{cal A} and $\mathcal{A}^2$ defined as
\begin{equation}
\mathcal{A}^2:= \mathcal{A}_+^2+\mathcal{A}_\times^2.
\end{equation}
It then follows from \eqref{Wightman lw regularized} and \eqref{Iku} that
\begin{eqnarray}
D^+_{\mathrm{lw}}(x-x')
&=&
    -i\int_{0}^{\infty}\frac{dk'_u}{2\pi}\, \frac{1}{2k'_u} \mathcal{I}(k'_u)\, e^{-i k'_u (u-u'-i \epsilon)}\nonumber\\
&=&
    -\frac{1}{8\pi^2}\frac{1}{\sqrt{1-\mathcal{A}^2}}
     \frac{1}{(u-u'-i \epsilon)(v-v'-i \epsilon)- R(\Delta x,\Delta y)/2}.
\end{eqnarray}
Finally, the result in the coordinates $(t,x,y,z)$ takes the form
\begin{equation}\label{D lw}
D^+_{\mathrm{lw}}(x-x')
=  -\frac{1}{4\pi^2}
    \frac{1}{\sqrt{1-\mathcal{A}^2}}\,
    \frac{1}{(t-t'-i\epsilon)^2 -(z-z')^2-R(\Delta x,\Delta y) }.
\end{equation}
When $\mathcal{A}_+=\mathcal{A}_\times=0$, \eqref{D lw} reduces to the ordinary Wightman function in the Minkowski spacetime with the standard $i\epsilon$-regularization:

\begin{equation}\label{D standard}
D_\mathrm{M}^+(x-x')
=  -\frac{1}{4\pi^2}\,
    \frac{1}{(t-t'-i\epsilon)^2 -\abs{\vec{x}-\vec{x}'}^2},
\end{equation}
where $\vec{x}\equiv(x,y,z)$.




\subsection{Short-wavelength limit}

Next, we consider the short-wavelength limit conditioned by \eqref{short-wavelength condition}. Neglecting any corrections in or higher than the order of $O(\Delta E/\omega)$, we take the formal limit $w\equiv\sqrt{2}\,\omega\rightarrow\infty$ upon \eqref{Wightman}. The result simply reduces to
\begin{equation}
D^+_{\mathrm{sw}}(x-x')\equiv \lim_{w\rightarrow \infty} D^+(x,x')
=   \int \frac{d^3k}{(2\pi)^3}\frac{1}{2k_u}
    e^{ i\bm{k}\cdot (\bm{x}-\bm{x}')}e^{-ik_u(u-u')-i\epsilon}e^{ -i\omega_{k_u,\bm{k}} (v-v-i\epsilon)},
\end{equation}
where the $i\epsilon$-regularization is again prescribed.
The dependence of the gravitational wave amplitude disappears, and $D_{\mathrm{sw}}(x-x')$ in the coordinates $(t,x,y,z)$ reads as
\begin{equation}\label{D sw}
D^+_{\mathrm{sw}}(x,x')=  -\frac{1}{4\pi^2}\,
    \frac{1}{    (t-t'-i\epsilon)^2
                 -\abs{\vec{x}-\vec{x}'}^2 },
\end{equation}
which formally is identical to the ordinary Wightman function in the Minkowski spacetime as given in \eqref{D standard}.

Although \eqref{D sw} apparently looks the same as that in the Minkowski spacetime, the physics it implies can be quite different from the latter. For one thing, the geodesic equation in the short-wavelength limit is different from that in the Minkowski spacetime (see \appref{app:geodesic eq}).

Once the detector's trajectory is given and the Wightman function is known, we are ready to compute the transition rate by \eqref{F dot}. First, we consider the case that the detector follows a free-falling (i.e.\ geodesic) trajectory, and then the case that the detector moves with a constant acceleration in the $z$ direction.

\section{Free-falling trajectory}\label{sec:free-falling}
In this section, we study the response of an Unruh-DeWitt detector falling freely in a gravitational wave background.
The free-falling trajectory is given by the geodesic equation, which is solved explicitly in \appref{app:geodesic eq}.

\subsection{Long-wavelength limit}
In the long-wavelength limit $\omega\rightarrow0$, according to \eqref{const velocity sol}, a free-falling trajectory takes the form
\begin{equation}\label{x freefall}
  x^\mu(\tau) = U^\mu\tau + x^\mu_0,
\end{equation}
where the 4-velocity
\begin{equation}
U^\mu=(u^t,u^x,u^y,u^z)
\end{equation}
is given by four constants, $u^t$, $u^x$, $u^y$, and $u^z$, subject to the constraint \eqref{velocity normalization 2}, and $x^\mu_0$ are displacement parameters.

Substituting \eqref{x freefall} into \eqref{D lw} yields
\begin{eqnarray}\label{D in freefall}
D^+_{\mathrm{lw}}(\tau,\tau')
&=& -\frac{1}{4\pi^2}\frac{1}{\sqrt{1-\mathcal{A}^2}}
     \frac{1}{(u_t \Delta \tau -i\epsilon)^2-(u_z \Delta\tau)^2+\frac{1-u_t^2 +u_z^2}{(1-\mathcal{A}^2)}\Delta\tau^2 - R(\Delta x,\Delta y)} , \nonumber\\
&=& -\frac{1}{4\pi^2}\frac{1}{\sqrt{1-\mathcal{A}^2}}
     \frac{1}{\Delta\tau^2((u^t)^2-(u^z)^2-i\epsilon')-\frac{\Delta\tau^2}{1-\mathcal{A}^2}((u^t)^2-(u^z)^2-1)}\nonumber\\
&=& -\frac{1}{4\pi^2}\frac{1-\mathcal{A}^2}{\sqrt{1-\mathcal{A}^2}}
     \frac{1}{\Delta\tau^2[(1-\mathcal{A}^2)((u^t)^2-(u^z)^2-i\epsilon')-((u^t)^2-(u^z)^2)+1]}\nonumber\\
&=& -\frac{1}{4\pi^2}\frac{1-\mathcal{A}^2}{\sqrt{1-\mathcal{A}^2}}
     \frac{1}{\Delta\tau^2[1-\mathcal{A}^2((u^t)^2-(u^z)^2)-i\epsilon'']}\nonumber\\
&=& -\frac{1}{4\pi^2}\frac{\sqrt{1-\mathcal{A}^2}}{1-\mathcal{A}^2((u^t)^2-(u^z)^2)}
     \frac{1}{(\Delta\tau^2-i\epsilon''')^2},
\end{eqnarray}
where $\epsilon$, $\epsilon'$, $\epsilon''$, and $\epsilon'''$ are infinitesimal positive numbers (rescaled differently).
The fact that $D^+_{\mathrm{lw}}(\tau,\tau')$ depends only on $\Delta\tau\equiv\tau-\tau'$ suggests that a free-falling Unruh-DeWitt detector is in equilibrium with $\phi$ in the long-wavelength limit.

Substituting \eqref{D in freefall} into \eqref{F dot} then yields the equilibrium transition rate
\begin{equation}
\dot{F}(\Delta E)
= - \frac{1}{4\pi^2} \frac{\sqrt{1-\mathcal{A}^2}}{1-\mathcal{A}^2((u^t)^2-(u^z)^2)}
    \int_{-\infty}^{\infty} d\Delta \tau\, \frac{e^{-i\Delta E \Delta\tau}}{(\Delta\tau-i\epsilon)^2}.
\end{equation}
If $\Delta \tau$ is considered to be a complex number, the transition rate can be calculated by a contour integral. The integrand has a pole of order 2 at $\Delta=i\epsilon$.

For $\Delta E>0$, the transition rate can be calculated by a contour integral along an infinite semicircle contour on the lower half of the $\Delta\tau$ plane. As the contour does not enclose the pole, the contour integral turns out to be zero.

For $\Delta E<0$, the integration can be calculated by a contour integral along an infinite semicircle contour on the upper half of the $\Delta\tau$ plane. The residue theorem applied to the pole at $\Delta\tau=i\epsilon$ gives
\begin{subequations}
\begin{eqnarray}
\dot{F}(\Delta E)
&=&- \frac{\Delta E}{2\pi}\frac{\sqrt{1-\mathcal{A}^2}}{1-\mathcal{A}^2((u^t)^2-(u^z)^2)}\\
&\equiv&- \frac{\Delta E}{2\pi}\frac{\sqrt{1-\mathcal{A}^2}}{1-\mathcal{A}^2(1+(u^x)^2+(u^y)^2 +\mathcal{A}_+\left((u^x)^2-(u^y)^2\right) +2\mathcal{A}_\times u^xu^y)},
\end{eqnarray}
\end{subequations}
where we have used \eqref{velocity normalization 2} to recast $(u^t)^2-(u^z)^2$ in terms of $u^x$ and $u^y$.\footnote{Under a rotation by $\theta$ around the $z$ axis, $u^{x}$, $u^y$, and $\mathcal{A}_{+/\times}$ transform as
\begin{eqnarray*}
  u^{x\prime} &=& \cos\theta\, u^x + \sin\theta\, u^y, \qquad
  u^{y\prime} = -\sin\theta\, u^x + \cos\theta\, u^y,\\
  \mathcal{A}'_+ &=&  \cos2\theta\mathcal{A}_+ + \sin2\theta\mathcal{A}_\times, \qquad
  \mathcal{A}'_\times = -\sin2\theta\mathcal{A}_+ + \cos2\theta\mathcal{A}_\times.
\end{eqnarray*}
Note that $\mathcal{A}^2$, $(u^x)^2+(u^y)^2$, and $\mathcal{A}_+\left((u^x)^2-(u^y)^2\right) +2\mathcal{A}_\times u^xu^y$ are all invariant under this transformation.}

In summary, we have
\begin{subequations}\label{F v}
\begin{eqnarray}
 \dot{F}(\Delta E)
 &=& 0, \quad \text{for}\ \Delta E > 0, \\
 &=&  - \frac{\Delta E}{2\pi} \frac{\sqrt{1-\mathcal{A}^2}}{1-\mathcal{A}^2((u^t)^2-(u^z)^2)}, \quad \text{for}\ \Delta E < 0.
\end{eqnarray}
\end{subequations}
As expected, in the limit that the gravitational wave amplitude goes to zero, i.e., $\mathcal{A}^2\rightarrow0$, \eqref{F v} reduces to the ordinary result in the Minkowski spacetime:
\begin{equation}\label{F v Minkowski}
\dot{F}(\Delta E)  \mathop{\longrightarrow}\limits_{\mathcal{A}^2\rightarrow0}
-\frac{\Delta E}{2\pi}\Theta(-\Delta E).
\end{equation}
Compared to the Minkowskian result, the transition rate \eqref{F v} is modified by an overall proportional factor that depends on the amplitude of the gravitational wave and the detector's velocity.
When $(u^t)^2 - (u^z)^2=1$ or, equivalently, $u^x=u^y=0$, \eqref{F v} in the case of $\Delta E < 0$ yields the maximum value:
\begin{equation}\label{F dot max}
\max_{U^\mu}\dot{F}(\Delta E)=- \frac{\Delta E}{2\pi} \frac{1}{\sqrt{1-\mathcal{A}^2}}.
\end{equation}
On the other hand, when $(u^t)^2 - (u^z)^2=0$ or, equivalently, $u^x$ and $u^y$ satisfy
\begin{equation}
1+(u^x)^2+(u^y)^2 +\mathcal{A}_+\left((u^x)^2-(u^y)^2\right) +2\mathcal{A}_\times u^xu^y=0,
\end{equation}
\eqref{F v} in the case of $\Delta E < 0$ yields the minimum value:
\begin{equation}
\min_{U^\mu}\dot{F}(\Delta E)=- \frac{\Delta E}{2\pi} \sqrt{1-\mathcal{A}^2}.
\end{equation}

It is instructive to compare \eqref{F v} with the case in flat spacetime with a compact
dimension. In a flat spacetime where the $z$ direction is compactified with a finite length $L$, the transition rate of the Unruh-DeWitt detector moving with the 4-velocity $U^\mu=(u^t,u^x,u^y,u^z)$ is given by (see \cite{Chiou:2016exd})
\begin{subequations}\label{rate for constant velocity}
\begin{eqnarray}
\label{rate for constant velocity a}
\dot{F}_L(\Delta E) &=& 0, \quad \text{for}\ \Delta E >0, \\
\label{rate for constant velocity b}
&=&
-\frac{\Delta E}{2\pi}
-\frac{i}{4\pi L u^t}
\ln
\left(
\frac{1-e^{i\frac{\Delta E L}{u^t+u^z}}}
{1-e^{-i\frac{\Delta E L}{u^t+u^z}}}
\,
\frac{1-e^{i\frac{\Delta E L}{u^t-u^z}}}
{1-e^{-i\frac{\Delta E L}{u^t-u^z}}}
\right),
\quad \text{for}\ \Delta E <0.
\end{eqnarray}
\end{subequations}
The correction due to the compact length $L$ in \eqref{rate for constant velocity} is scaled as $O(L^{-1})$, which vanishes in the formal limit $L\rightarrow\infty$. By contrast, the leading-order correction due to the gravitational wave is scaled as $O((1/\omega)^0)$ as shown in \eqref{F v}, which survives the formal limit $\omega\rightarrow0$, i.e., as the gravitational wavelength goes to infinity.
In both cases, the transition rates are different from the ordinary result in the Minkowski spacetime, essentially because the mode expansion of the quantum field $\phi$ is altered in the presence of the compact dimension or the gravitational wave. However, the gravitational wave effect is more involved and cannot be explained out simply by saying that the gravitational wave imposes a large length scale of the wavelength $1/\omega$ as well as the compact dimension does of the finite length $L$.

It is rather surprising that a detector that apparently has no spatial extent can sense the presence of a gravitational wave, as sensing tidal force requires a certain spatial extent. One might try to argue that the Unruh-DeWitt detector has an intrinsic energy scale $\Delta E$ and therefore, according to the energy-time uncertainty principle, exhibits a temporal scale $\sim1/\Delta E$, which in turn gives rise to a spatial scale $\sim v/\Delta E$, where $v<1$ is the velocity of the detector. However, even if the detector has a finite extent $\delta\ell$ (and even if $\delta\ell$ is much larger than ${\sim}1/\Delta E$ for whatever reason),\footnote{In fact, particle detectors with finite spatial extent have been discussed in the literature \cite{Grove:1983rp}.} the tidal force produced by a gravitational wave over a spatial separation of $\delta\ell$ is proportional to $(\delta\ell)\ddot{h}_{+/\times}\sim \mathcal{A}_{+/\times}\omega^2\delta\ell$, which vanishes in the limit $\omega\rightarrow0$. Since the correction in \eqref{F v} survives the formal limit $\omega\rightarrow0$, this gravitational wave effect on a quantum system is qualitatively different from that on a classical mechanical system, and cannot be understood in terms of gravitational wave tidal force. This is a genuine quantum effect that has no classical analogue.

\subsection{Short-wavelength limit}
The solution to the geodesic equation in a gravitational wave background is given by \eqref{geodesic eq sol}, which in general is very complicated.
The solution takes a simple form in the long-wavelength limit as given by \eqref{const velocity sol}, but it remains complicated under the short-wavelength condition \eqref{short-wavelength condition}.

Therefore, even though the Wightman function in the limit $\omega\rightarrow\infty$ as given in \eqref{D sw} apparently is identical to the ordinary Wightman function in the Minkowski spacetime, a free-falling Unruh-DeWitt detector in a gravitational wave background in general is \emph{not} in equilibrium  with $\phi$, contrary to that in the Minkowski spacetime.

However, we do have a special geodesic solution given by \eqref{trivial sol}, which is simple and corresponds to a free-falling trajectory moving in the propagation direction of the gravitational wave, i.e.,
\begin{equation}\label{trivial sol x}
x^\mu(\tau) = (u^t\tau,0,0,u^z\tau) + x^\mu_0.
\end{equation}
Substituting this trajectory into $D^+_{\mathrm{sw}}(x,x')$ in \eqref{D sw}, we see that $D^+_{\mathrm{sw}}(\tau,\tau')$ depends only on $\Delta\tau\equiv\tau-\tau'$. Therefore, the Unruh-DeWitt detector that freely falls along the trajectory \eqref{trivial sol x} is in equilibrium with $\phi$ in the short-wavelength limit $\omega\rightarrow\infty$.
Since $D^+_{\mathrm{sw}}(x,x')$ is formally the same as the ordinary Wightman function in the Minkowski spacetime, the transition rate $\dot{F}(\Delta)$ along \eqref{trivial sol x} is the same as the ordinary result in the Minkowski spacetime, signaling no presence of the gravitational wave at all. This can be understood intuitively: since the oscillation of the gravitational wave is much faster than the detector's response time ${\sim}1/\Delta E$ for the two-level transition, the detector has no time to respond to the driving oscillation (provided that the gravitational wave is weak enough so that the linearized theory is legitimate).

\section{Constant-accelerating trajectory}\label{sec:constant-accelerating}
In this section, we study the response of an Unruh-DeWitt detector that moves with a constant acceleration $1/\alpha$ in the $z$ direction. The trajectory is given by
\begin{equation}\label{x a}
t= \alpha \sinh{\frac{\tau}{\alpha}}+t_0,\quad x= y=\mathrm{const},\quad z= \alpha\cosh{\frac{\tau}{\alpha}}+z_0,
\end{equation}
where $t_0$ and $z_0$ are displacement parameters.

\subsection{Long-wavelength limit}
In the long-wavelength limit, substituting \eqref{x a} into $D^+_\mathrm{lw}(x,x')$ in \eqref{D lw}, we have\footnote{The derivation involves some details, which can be found in Appendix C of \cite{Chiou:2016exd}.}
\begin{eqnarray}\label{D in acceleration}
D^+_\mathrm{lw}(\Delta \tau)
&=& -\frac{\alpha^2}{16\pi^2} \frac{1}{\sqrt{1-\mathcal{A}^2}}\frac{1}{\sinh^2{(\frac{\Delta\tau}{2\alpha}-\frac{i\epsilon}{2\alpha})}} \nonumber\\
&=& -\frac{1}{4\pi^2} \frac{1}{\sqrt{1-\mathcal{A}^2}}
\sum_{k=-\infty}^{\infty} \frac{1}{(\Delta \tau-i\epsilon+2\pi ik\alpha)^2},
\end{eqnarray}
where we have applied the identity
\begin{equation}
  \csc^2{\pi x}=\frac{1}{\pi^2}\sum_{k=-\infty}^{\infty}\frac{1}{(x-k)^2}.
\end{equation}
As $D^+_{\mathrm{lw}}(\tau,\tau')$ depends only on $\Delta\tau\equiv\tau-\tau'$, the Unruh-DeWitt detector is in equilibrium with $\phi$.

Substituting \eqref{D in acceleration} into \eqref{F dot} and performing the contour integral, we obtain the transition rate
\begin{equation}\label{F a}
\dot{F}(\Delta E)= \frac{\Delta E}{2\pi}\,\frac{1}{\sqrt{1-\mathcal{A}^2}}\, \frac{1}{e^{2\pi\Delta E\alpha}-1},
\end{equation}
for both $\Delta E>0$ and $\Delta E<0$.
Except for the overall proportional factor $(1-\mathcal{A}^2)^{-1/2}$, this result is exactly the same as the ordinary result of a constant-accelerating Unruh-DeWitt detector moving in the Minkowski spacetime given by\footnote{It is often said that the transition rate \eqref{F a Minkowski} for a constant-accelerating detector moving in the Minkowski spacetime corresponds to the transition rate for a detector lying at rest in a thermal bath of particles of $\phi$ at the Unruh temperature $T=(2\pi k_\mathrm{B}\alpha)^{-1}$. This is common confusion, and it is only a coincidence solely for the case of $m=0$ and $\Delta E>0$ that the transition rate of the former happens to be identical to that of the latter. (See Sec.\ III.A.4 of \cite{Crispino:2007eb} for a detailed clarification for this confusion.) In fact, as discussed in \secref{sec:Unruh-DeWitt detector} (and more in Appendix~A of \cite{Chiou:2016exd}), the temperature of a detector, including the Unruh temperature \eqref{Unruh temperature}, is a notion of detailed balance between a transition process and its reverse process via \eqref{detailed balance}, whereas the transition rate \textit{per se} does not makes any sense of temperature.}
\begin{equation}\label{F a Minkowski}
\dot{F}(\Delta E)= \frac{\Delta E}{2\pi}\, \frac{1}{e^{2\pi\Delta E\alpha}-1}.
\end{equation}
The overall proportional factor is in perfect agreement with the free-falling case with $u^x=u^y=0$ as shown in \eqref{F dot max}.

It is again instructive to compare \eqref{F a} with the case in flat spacetime with a compact
dimension. In a flat spacetime where one spatial dimension perpendicular to the $z$ direction is compactified with a finite length $L$, the transition rate of the Unruh-DeWitt detector moving along \eqref{x a} is given by (see \cite{Chiou:2016exd})
\begin{equation}\label{rate for a in z}
\dot{F}_L(\Delta E)
= \frac{\Delta E}{2\pi} \frac{1}{e^{2\pi\Delta E\alpha}-1}
- \Theta(-\Delta E) \sum_{n=1}^\infty
\frac{\sin\left(2\alpha\Delta E\sinh^{-1}\frac{nL}{2\alpha}\right)}
{n\pi L\sqrt{1+\left(\frac{nL}{2\alpha}\right)^2}}.
\end{equation}
The correction due to the compact length $L$ in \eqref{rate for a in z} is scaled as $O(L^{-1})$, which vanishes in the formal limit $L\rightarrow\infty$. By contrast, the leading-order correction due to the gravitational wave is scaled as $O((1/\omega)^0)$ as shown in \eqref{F a}, which survives the formal limit $\omega\rightarrow0$.

Just as we have commented for the case of a free-falling detector, the gravitational wave effect on a constant-accelerating Unruh-DeWitt detector is a genuine quantum effect that cannot be understood in terms of gravitational wave tidal force.

\subsection{Short-wavelength limit}
In the short-wavelength limit, substituting the constant-accelerating trajectory \eqref{x a} into $D^+_\mathrm{sw}(x,x')$ in \eqref{D sw} yields the same result in the Minkowski spacetime.
Consequently, in the short-wavelength limit $\omega\rightarrow\infty$, the detector moving along \eqref{x a} is in equilibrium with $\phi$, and the transition rate is
the same as the ordinary result in the Minkowski spacetime.

Again, this can be understood intuitively: the gravitational oscillates so fast that the detector has no time to respond to it.

\section{Summary and remarks}\label{sec:summary}
Applying the techniques of light-front quantization used in the literature of QCD, we have successfully quantized the real scalar field $\phi(x)$ in a monochromatic gravitational wave background, obtaining the formulae \eqref{a and a dag} and \eqref{phi}. This enables us to compute the corresponding Wightman function $D^+(x,x')$ as given by \eqref{Wightman}, which is greatly simplified in the long-wavelength limit $\omega\rightarrow0$ and the short-wavelength limit $\omega\rightarrow\infty$, as given by \eqref{D lw} and \eqref{D sw}, respectively.

With the Wightman function at hand, we then investigate the response of the Unruh-DeWitt detector in a gravitational wave background for the two cases of a free-falling trajectory and a constant-accelerating trajectory moving in the propagation direction of the gravitational wave.

In the long-wavelength limit $\omega\rightarrow0$, the equilibrium transition rate of the detector moving along a free-falling trajectory \eqref{x freefall} is given by \eqref{F v}, and that along a constant-accelerating trajectory \eqref{x a} is given by \eqref{F a}.
These results are different from their corresponding counterparts \eqref{F v Minkowski} and \eqref{F a Minkowski}, respectively, in flat spacetime (without any gravitational wave) by an overall proportional factor, which depends on the amplitude of the gravitational wave but not the gravitational wave wavelength $1/\omega$. That is, in both cases, the leading-order correction due to the gravitational wave is of $O((\omega\Delta)^0)$, which survives the formal limit $\omega\rightarrow0$ as long as the gravitational wave amplitude remains finite. This suggests that the gravitational wave effect on the Unruh-DeWitt detector is more involved than merely imposing a large length scale of the wavelength as well as the presence of spatial boundaries does of the length scale delimited by the boundaries (see \cite{Chiou:2016exd,Davies:1989me}).
Furthermore, even if we suppose that the Unruh-DeWitt detector has a finite spatial extent, this effect is qualitatively different from that on a classical mechanical system and cannot be explained out in terms of gravitational wave tidal force. This is a genuine quantum effect that has no classical analogue.

On the other hand, in the short-wavelength limit $\omega\rightarrow\infty$, the Unruh-DeWitt detector following a free-falling trajectory in a gravitational wave background is \emph{not} in equilibrium with the field $\phi$ in general, except for the free-falling trajectory moving in the propagation direction of the gravitational wave as given by \eqref{trivial sol x}. The equilibrium transition rate along \eqref{trivial sol x} is the same as the ordinary result in flat spacetime, showing no response to the gravitational wave.
Furthermore, along a constant-accelerating trajectory given by \eqref{x a}, the equilibrium transition rate is again the same as the ordinary result in flat spacetime.
The fact that the Unruh-DeWitt detector in equilibrium with $\phi$ does not respond to the gravitational wave background in the limit $\omega\rightarrow\infty$ can be understood intuitively: the gravitational wave oscillates so fast that the detector has no time to respond to the driving oscillation within the timescale ${\sim}1/\Delta E$ for the two-level transition.

The results of our study also raise some open questions. We have demonstrated that the transition rate of an Unruh-DeWitt detector can be affected by the presence of a gravitational wave. It is unclear whether the gravitational wave is involved with energy transfer for the transition process \eqref{transition} as it is for the response of a classical mechanical system, or perhaps it merely acts as a ``catalyst'', which increases the transition efficiency but does not deposit or withdraw any net energy.
Neither does our study investigate the aspect of detailed balance. It is uncertain whether detailed balance can be established in some particular settings in a gravitational wave background. If detailed balance can be established after all, it is important to know whether the temperature of detailed balance is shifted by the gravitational wave background and whether the temperature shift can be understood in terms of energy transfer from the gravitational wave.
Furthermore, it was recently shown that the concurrence of transition probability of a pair of free-falling Unruh-DeWitt detectors, which serves as a probe of vacuum entanglement, responds to the presence of a gravitational wave and exhibits certain resonance effects \cite{Xu:2020pbj}. As our study considers arbitrary gravitational wave polarization and more general trajectories, including both free-falling and constant-accelerating ones, our results may help to investigate the interplay between vacuum entanglement and gravitational waves in broader settings.

The analysis in this paper is performed entirely in the framework of quantum field theory in curved spacetime. The scalar field $\phi$ is quantized in the gravitational wave background, and the Unruh-DeWitt detector is modeled as a quantum system coupled to $\phi$. The gravitational field, on the other hand, is treated completely as a classical background and not quantized at all. Although the gravitational wave effect on the Unruh-DeWitt detector is a quantum effect with no classical analogue, it is not a consequence of quantum gravity. Nevertheless, it is an intriguing open question whether the effect we found here can be understood in terms of a quantum detector coupled to both particles of $\phi$ and gravitons (quantized particles of the degrees of freedom of gravitational waves). This question might also be related to the aforementioned issue of energy transfer from the gravitational wave.

Finally, while it is conceptually important to understand the effects of a gravitational wave on a quantum system, it should be remarked that our investigation on the Unruh-DeWitt in response to a gravitational wave is mainly for theoretical concerns. Experimentally, measuring the response of the Unruh-DeWitt detector is extremely challenging, if not completely out of reach of current technology. In the case of a constant-accelerating trajectory, an experimentally reachable value for the acceleration $1/\alpha$ is extremely small compared to $\abs{\Delta E}$ (i.e., $\abs{\Delta E}\alpha\gg1$) for a typical two-level quantum system of which one can reliably measure the transition rate. This renders the transition rate given by \eqref{F a} experimentally indistinguishable from the result of a free-falling trajectory as given by \eqref{F v}.
With regard to the transition rate \eqref{F v} of a free-falling trajectory, its response to the gravitational wave seems to be measurable for $\Delta E<0$, provided that the amplitude of the gravitational wave is strong enough. However, as the transition rate, in principle, has to be measured by a large ensemble of identical Unruh-DeWitt detectors (see \secref{sec:Unruh-DeWitt detector} and Appendix A of \cite{Chiou:2016exd} for more discussions), the interactions between detectors of the ensemble and between the system and its surroundings will inevitably introduce noises that will spoil the signal in response to a gravitational wave that is extremely weak when arriving on earth.

Nevertheless, as the techniques of quantum measurement advance drastically in recent years, it might be possible to overcome the noise problem and eventually use an Unruh-DeWitt-type quantum system as a gravitational wave detector in the near future. Compared to resonance mass detectors and interferometric gravitational wave detectors, a quantum detector is expected to be sensitive to gravitational waves of much higher frequencies, since the characteristic timescale of a quantum system is typically much shorter than that of a resonance mass detector or an interferometric gravitational detector.
Furthermore, an Unruh-DeWitt-type detector may also have a quite wide bandwidth of sensitivity, because, as indicated by \eqref{F v}, the leading-order correction is insensitive to $\omega$ in the long-wavelength limit.
In order for the Unruh-DeWitt-type detector to be used as a gravitational wave detector, for the theoretical aspect, one will have to perform a detailed numerical analysis to know how exactly it responds to any arbitrary wavelength $1/\omega$, not only the results obtained in this paper for the long-wavelength and short-wavelength extremes.


\begin{acknowledgments}
The authors would like to thank an anonymous reviewer of the previous manuscript for raising some important issues, which have helped to improve this paper significantly. This work was supported in part by the Ministry of Science and Technology, Taiwan under the Grants No.\ 110-2112-M-002-016-MY3 and No.\ 110-2112-M-110-015.
\end{acknowledgments}


\appendix

\section{Geodesic equation in a gravitational wave background}\label{app:geodesic eq}
In this appendix, we solve the geodesic equation in a monochromatic gravitational wave background of an arbitrary elliptical polarization (linear and circular polarizations are special cases).

In the light-front coordinates $(u,v,x,y)$, the metric of a monochromatic gravitational background wave propagating in the $z$ direction is given by
\begin{equation}\label{g mu nu}
g_{\mu\nu}
=\left(\begin{array}{cccc}
         0 & -1 & 0 & 0 \\
         -1 & 0 & 0 & 0 \\
         0 & 0 & 1+h_+(u) & h_\times(u) \\
         0 & 0 & h_\times(u) & 1-h_+(u)
       \end{array}\right),
\end{equation}
where $h_{+/\times}(u):=A_{+/\times}\cos(\omega u+\theta_{+/\times})$ in accord with \eqref{h}.
The lower-indexed Christoffel connection is defined as
\begin{equation}
\Gamma_{\mu\nu\alpha}
:=
g_{\alpha\beta}\Gamma^{\beta}_{\mu\nu}
\approx
\frac{1}{2}
(h_{\alpha\nu,\mu}+h_{\mu\alpha,\nu}-h_{\mu\nu,\alpha}).
\end{equation}
In the coordinates $(u,v,x,y)$, the components of $\Gamma_{\mu\nu\alpha}$ are given by
\begin{subequations}
\begin{eqnarray}
\Gamma_{xxu}&=&-\Gamma_{xux}=-\Gamma_{yyu}=\Gamma_{yuy}=\frac{-h'_+(u)}{2}= \frac{A_{+}\omega}{2} \sin(\omega u+\theta_+),\\
\Gamma_{xyu}&=&-\Gamma_{xuy}=-\Gamma_{yux}=\frac{-h'_\times(u)}{2}=\frac{A_{\times}\omega}{2} \sin(\omega u+\theta_\times),
\end{eqnarray}
\end{subequations}
where $h'_{+/\times}:=\partial_u h_{+/\times}$, and all the other components vanish.
These lead to
\begin{subequations}
\begin{eqnarray}
\Gamma^{v}_{xx} &=& -\Gamma^{v}_{yy} = \Gamma^{x}_{ux}  =-\Gamma^{y}_{uy}  = \frac{h'_+}{2}, \\
\Gamma^{v}_{xy} &=&  \Gamma^{y}_{ux} = \Gamma^{x}_{uy} = \frac{h'_\times}{2}.
\end{eqnarray}
\end{subequations}
Consequently, the geodesic equation is given by
\begin{subequations}\label{geodesic eq}
\begin{eqnarray}
\frac{dU^u}{d\tau}
&=& 0, \\
\frac{dU^v}{d\tau}
&=& -\Gamma^{v}_{xx} U^xU^x- \Gamma^{v}_{yy}  U^yU^y     \\\nonumber
&=&  \frac{h'_{+}}{2}(U^yU^y- U^xU^x)- h'_\times U^xU^y, \\
\frac{dU^x}{d\tau}
&=& -2 \Gamma^{x}_{ux}  U^uU^x-2 \Gamma^{x}_{uy}  U^uU^y \\\nonumber
&=& -(h'_+\, U^x - h'_\times\, U^y)\,U^u, \\
\frac{dU^y}{d\tau}
&=& -2 \Gamma^{y}_{uy}  U^uU^y -2 \Gamma^{y}_{ux}  U^uU^x \\\nonumber
&=&( h'_+\,U^y - h'_\times\, U^x)\,U^u ,
\end{eqnarray}
\end{subequations}
where $U^\mu$ is the 4-velocity and is subject to the constraint
\begin{equation}\label{velocity normalization}
  g_{\mu\nu}U^\mu U^\nu= -1.
\end{equation}

The solution of the differential equation \eqref{geodesic eq} is given by
\begin{subequations}\label{geodesic eq sol}
\begin{eqnarray}
U^u
&=&  \text{constant}, \\
\label{geodesic eq sol Uv}
U^v
&=& c_3+\int \frac{du}{U^u}\left(\frac{h'_{+}}{2}(U^yU^y- U^xU^x)- h'_\times U^xU^y \right),\\
\left(\begin{array}{c}
        U^x \\
        U^y
      \end{array}\right)
&=&c_1 \,\mathbf{a}_1(u)\, e^{\lambda(u)}+ c_2\, \mathbf{a}_2(u)\, e^{-\lambda(u)},
\end{eqnarray}
\end{subequations}
where
\begin{equation}
u= U^u \tau,
\end{equation}
and where the function $\lambda(u)$ is
\begin{eqnarray}
\lambda(u)
&=& \int du \lambda'=\int du \sqrt{{h'}_{+}^{2}+{h'}_{\times}^{2}}\\
&=& \omega\int du \sqrt{ A^2_{+}\sin^2(\omega u+\theta_+)^2 +A^2_{\times}\sin^2(\omega u+\theta_\times)},
\end{eqnarray}
the doublets $\mathbf{a}_1(u)$, $\mathbf{a}_2(u)$ are solutions to
\begin{equation}
\left(\begin{array}{cc}
        -h'_+ \mp \lambda' & -h'_\times \\
        -h'_\times & h'_+\mp \lambda'
      \end{array}\right) \mathbf{a}_{1,2} = 0,
\end{equation}
and $c_1$, $c_2$, and $c_3$ are constants to be determined by the initial condition.

Particularly, in the case of a linear polarization in the $+$ mode, by setting $A_\times=0$ we have the closed-form solution given by
\begin{eqnarray}
U^u &=& \text{constant}, \\
U^x
&=& c_1\, e^{-A_{+}\cos(\omega u)}, \\
U^y
&=& c_2\, e^{A_{+}\cos(\omega u)}, \\
U^v
&=& \frac{-A_{+}\omega}{2}
    \int \frac{du}{U^u} \sin(\omega u)
                \left(      c_2^2 e^{2A_{+}\cos(\omega u)}
                          - c_1^2 e^{-2A_{+}\cos(\omega u)}
                          \right) \nonumber\\
&=& c_3+\frac{1}{4U^u} \left(c_1^2\,  e^{-2A_{+}\cos(\omega u)}+c_2^2\, e^{2A_{+}\cos(\omega u)}\right).
\end{eqnarray}
The closed-from solution for the $\times$ mode is similar.

The general solution \eqref{geodesic eq sol} admits a simple solution in the case of $c_1=c_2=0$.
Taking $c_1=c_2=0$ corresponds to $U^x=U^y=0$. Consequently, the integrand in \eqref{geodesic eq sol Uv} is zero, and thus $U^v$ is a constant of motion as well as $U^u$. In the coordinates $(t,x,y,z)$, we then have
\begin{equation}\label{trivial sol}
U^\mu=(u^t,0,0,u^z),
\end{equation}
where $u^t$ and $u^t$ are constants subject to \eqref{velocity normalization}. That is, if a free-falling point object moves in the $z$ direction, its geodesic trajectory looks as if the gravitational wave were absent. This is anticipated, as the geodesic deviation induced by the gravitational wave is only in the transverse (i.e. $x$ and $y$) directions.

In the long-wavelength limit $\omega\rightarrow0$, we have $h'=-\omega\sin(\omega u+\theta_{+/\times})\rightarrow 0$, and thus dependence on $u$ becomes negligible in \eqref{geodesic eq sol}. Consequently, $U^u$, $U^v$, $U^x$, and $U^y$ are all constants of motion.
In the coordinates $(t,x,y,z)$, we then have
\begin{equation}\label{const velocity sol}
U^\mu = (u^t,u^x,u^y,u^z),
\end{equation}
where $u_t$, $u_x$, $u_y$, and $u_z$ are all constants subject to \eqref{velocity normalization}.
In the limit $\omega\rightarrow0$, the metric \eqref{g mu nu} in the coordinates $(t,x,y,z)$ takes the form
\begin{equation}
g_{\mu\nu}
\mathop{\longrightarrow}\limits_{\omega \rightarrow0}
\left(\begin{array}{cccc}
         -1 & 0 & 0 & 0 \\
         0 & 1 & 0 & 0 \\
         0 & 0 & 1+\mathcal{A}_+ & \mathcal{A}_\times \\
         0 & 0 & \mathcal{A}_\times & 1-\mathcal{A}_+
       \end{array}\right),
\end{equation}
where $\mathcal{A}_{+/\times}$ are defined in \eqref{cal A}.
The condition \eqref{velocity normalization} then reads as
\begin{equation}\label{velocity normalization 2}
1+(u^x)^2+(u^y)^2 +\mathcal{A}_+\left((u^x)^2-(u^y)^2\right) +2\mathcal{A}_\times u^xu^y = (u^t)^2-(u^z)^2.
\end{equation}



\end{document}